\documentclass[useAMS,usenatbib]{mn2e}

\usepackage{graphicx}
\usepackage{caption}
\usepackage{times}
\usepackage{natbib}
\usepackage{graphics}
\usepackage{epsfig}
\usepackage{array}
\usepackage{stfloats}
\usepackage{fixltx2e}
\usepackage{amsmath}
\usepackage[lofdepth,lotdepth]{subfig}
\usepackage[a4paper,colorlinks=true,pdfstartview=FitV,linkcolor=red,citecolor=blue,urlcolor=magenta]{hyperref}
\usepackage{float}

\newcommand{\gtrsim}{\ga}

\newcommand{\gsim}{\,\lower2truept\hbox{${>\atop\hbox{\raise4truept\hbox{$\sim$}}}$}\,}

\newcommand{\be}{\begin{equation}}
\newcommand{\ee}{\end{equation}}
\newcommand{\bea}{\begin{eqnarray}}
\newcommand{\eea}{\end{eqnarray}}

\renewcommand{\vec}[1]{ {\bmath #1} } 

\def\ltsima{$\; \buildrel < \over \sim \;$}
\def\simlt{\lower.5ex\hbox{\ltsima}}
\def\gtsima{$\; \buildrel > \over \sim \;$}
\def\simgt{\lower.5ex\hbox{\gtsima}}
\newcommand{\Mpch}{$h^{-1}\,\mbox{Mpc}$\,}
\newcommand{\Gpch}{$h^{-1}\,\mbox{Gpc}$\,}

\title[Cosmic voids in the CoDECS simulations]{Cosmic voids in coupled dark energy cosmologies: the impact of halo bias}
\author[G.~Pollina et al.]{\parbox{\textwidth}{Giorgia Pollina$^{1,4,5}$, Marco Baldi$^{1,2,3}$, Federico Marulli$^{1,2,3}$, Lauro Moscardini$^{1,2,3}$}
\\
\\$^{1}$Dipartimento di Fisica e Astronomia, Alma Mater Studiorum Universit\`a di Bologna, viale Berti Pichat, 6/2, I-40127 Bologna, Italy
\\$^{2}$INAF - Osservatorio Astronomico di Bologna, via Ranzani 1, I-40127 Bologna, Italy
\\$^{3}$INFN - Sezione di Bologna, viale Berti Pichat 6/2, I-40127 Bologna, Italy
\\$^{4}$Universit\"ats-Sternwarte M\"unchen, Fakult\"at f\"ur Physik, Ludwig-Maximilians Universit\"at M\"unchen, Scheinerstr. 1, D-81679 M\"unchen, Germany
\\$^{5}$Excellence Cluster Universe,  Boltzmannstr. 2, D-85748 Garching, Germany}

\begin{document}
%\date{Accepted ???. Received ???; in original form }
\pagerange{\pageref{firstpage}--\pageref{lastpage}} \pubyear{2011}
\maketitle
\label{firstpage}
\begin{abstract}	
  In this work we analyse the properties of cosmic voids in standard
  and coupled dark energy cosmologies. Using large numerical
  simulations, we investigate the effects produced by the dark energy
  coupling on three statistics: the filling factor, the size
  distribution and the stacked profiles of cosmic voids.  We find that
  the bias of the tracers of the density field used to identify the
  voids strongly influences the properties of the void
  catalogues, and, consequently, the possibility of using the
  identified voids as a probe to distinguish coupled dark energy
  models from the standard $\Lambda $CDM cosmology. In fact, on one
  hand coupled dark energy models are characterised by an excess of
  large voids in the cold dark matter distribution as compared to the
  reference standard cosmology, due to their higher normalisation of
  linear perturbations at low redshifts. Specifically, these models
  present an excess of large voids with $R_{\rm eff}>20, 15, 12$
  \Mpch, at $z=0, 0.55, 1$, respectively.  On the other hand, we do
  not find any significant difference in the properties of the voids
  detected in the distribution of collapsed dark matter halos. These
  results imply that the tracer bias has a significant impact on the
  possibility of using cosmic void catalogues to probe cosmology.
\end{abstract}


%*****************************************************************************

\section{Introduction}
\label{i}

Despite the fact that the presently accepted standard cosmological
model, the so-called $\Lambda$CDM scenario, appears to be fully
consistent with most of the available observations \citep[see
e.g.][]{2015Planck}, it still presents some open issues in the
detailed description of the distribution of matter at small scales.
%{ 
%{\citep[see e.g.][]{Moore_1994,KuziodeNaray_etal_2006,Oh_etal_2011,Salucci_etal_2012,BoylanKolchin_Bullock_Kaplinghat_2011,BoylanKolchin_Bullock_Kaplinghat_2012}}  as well as in some statistical
%properties of the large-scale structures {\citep[se
%  e.g.][]{Planck_XX,Planck_XXIV}}. }
One of such properties that still appears problematic is the observed
abundance of dwarf galaxies in the underdense regions of the Universe,
which is found to be significantly lower than what predicted by large
N-body simulations carried out within the $\Lambda $CDM cosmology.
This problem, that was pointed out for the first time by
\citet{peebles2001}, goes under the name of the {\em void phenomenon},
and it has been discussed by several authors over the past years
\citep[see e.g.][]{Tinker2009, sutter2014cDE}.

Besides the poor theoretical understanding of a cosmological constant
as source of the observed accelerated expansion of the Universe
\citep[][]{weinberg1989cosmological}, the void phenomenon is therefore
one of the few observational tensions that motivate the investigation
of alternative cosmological scenarios, together with the so-called
{\em cusp-core problem} \citep[][]{deBlok2010}, the {\em satellite
  problem} \citep[][]{Bullok2010}, the {\em too big to fail problem}
\citep[][]{Boylan2011}, and the recently detected tension between the
CMB- and cluster-based estimations of $\sigma _{8}$, the r.m.s. of the
mass density field within a sphere of radius 8 \Mpch
\citep[][]{Planck2015param}.

A relevant class of alternative cosmological models that has been
widely investigated in recent years is given by the so-called coupled
dark energy scenario \citep[cDE hereafter, see
e.g.][]{Wetterich_1995,Amendola_2000,Amendola_2004,Farrar2004,Baldi_2011a}.
In these models a dynamical scalar field sourcing the accelerated
cosmic expansion \citep[see e.g.][]{wetterich1988, RatraPeebles1988}
is coupled to cold dark matter (CDM) particles resulting in a
direct exchange of energy-momentum between these two cosmic
components.  Such interaction gives rise to a new fifth force acting
on CDM particles, possibly capable to make the voids emptier
\citep{nusser2005}.  Other possible ways to address the void
phenomenon have been proposed, such as, for example, a modification of
gravity at very large scales \citep{LiZhao2009, clampitt2013,
  spolyar2013}.

The main effects of cDE models on the large-scale matter distribution
in the Universe, as well as on the structural properties of highly
nonlinear collapsed objects (such as galaxies and galaxy clusters),
have been widely investigated in the recent past by several works
mostly based on dedicated large N-body simulations \citep[see
e.g.][]{Maccio_etal_2004, Baldi_etal_2010, Li_Barrow_2011,
  Baldi_2011b, CoDECS, vera2012, Marulli_Baldi_Moscardini_2012,
  Giocoli_etal_2012, Moresco_etal_2014, Carlesi_etal_2014a,
  Carlesi_etal_2014b}.  However, in these rich and high-dense
environments the effects produced by cDE are expected to be
significantly modified by the complex and not yet fully understood
baryonic processes occurring within astrophysical objects.  Therefore,
studying the properties of underdense regions of the universe might
represent a complementary approach to investigate cDE scenarios, and
might provide a direct test of such cosmological models through a
direct comparison with the properties of the observed cosmic voids
(hereafter CV).

Although the existence of CV - defined as large
underdense regions of the Universe - was one of the earliest
predictions of the standard cosmological model \citep{hausman1983},
and the observational discovery of CV dates back to over than 30 years
ago \citep[see][]{GregoryThompson1978, kirshner1981}, it is only in
recent years that systematic studies about CV have become possible
thanks to the increasing depth and volume of current galaxy surveys
and to the advent of large numerical simulations that allow to predict
with high accuracy the topology of the cosmic web.

The recent interest for CV is mostly related to their yet unexploited
potential to probe cosmological models and constrain cosmological
parameters, thanks to the claimed universality of their general
statistical and structural properties \citep[see e.g.][]{colberg2005,
  ricciardelli2013, ricciardelli2014}. In particular, CV might
represent a population of ideal spheres with a homogeneous
distribution in the Universe at different redshifts, so that their
size evolution can be used to characterise the expansion history of
the Universe by means of the Alcock \& Paczynski (AP) test
\citep{AlcockPaczynski1979}, as already pointed out by recent works
\citep{sutter2012, sutter2014}.

Furthermore, CV might have an impact on the observed properties of the
Cosmic Microwave Background (CMB), and the persisting CMB anomalies
like the Cold Spot could be explained as resulting from the Integrated
Sachs-Wolfe effect over large CV, as suggested by different works \citep{rees1968,
  szapudi2014, kovac2014, Finelli_etal_2014}.  The next generation of
large galaxies surveys such as the ESA Euclid mission \citep{laureijs2011,
  amendola2013} are expected to detect gravitational lensing from
medium size CV with which it will be possible to directly constrain
the void density profiles without resorting on luminous tracers like
galaxies, which would require to model their bias \citep{izumi2013,
  krause2013, melchior2014, clampitt2014}.

CV are therefore one of the most appealing and promising cosmological
probes: being almost empty, their growth during the cosmic history
should be at most weakly nonlinear and their properties could be
possibly affected by the nature of DE and by the properties of the
primordial density field in which they evolve \citep{odrzywolek2009,
  damico2011,bos2012,gibbons2014}.  In particular, the shape of CV has been
shown to be very sensitive to the equation of state of the DE
component \citep[][]{lavaux2010}.  Defining the properties of CV in
different cosmological models can then represent an important handle
to discriminate between these models.

The present work focuses on the investigation of the properties of CV
in the standard $\Lambda$CDM cosmology, as well as in a series of
competing cDE models. This has been done by extracting the population
of CV from both the cold dark matter (CDM) and the halo distributions
arising in large cosmological N-body simulations of these different
cosmological scenarios. To this end, we made use of the publicly
available data of the {\small CoDECS} simulations
\citep[][]{baldi2012}, including three different models of DE
interaction besides a $\Lambda$CDM reference run.  We identified CV in
the {\small CoDECS} runs with {\small VIDE} \citep[Void IDentification
and Examination toolkit,][]{VIDE2015}, a substantially modified
version of the publicly-available void finder {\small ZOBOV}
\citep[ZOnes Bordering On Voidness,][]{neyrinck2008}, and compared the
statistical and structural properties of the resulting void catalogs.
Our results show that cDE models are characterised by an excess of
large CV in the CDM distribution with respect to the reference
$\Lambda $CDM cosmology, as expected from their higher normalisation
of linear perturbations at low redshifts.  This is consistent with the
theoretical predictions on the abundance of CV presented in
\citet{Pisani_etal_2015}, while the latter work seems to be in
contrast with the recent findings of \citet{Sutter_etal_2015} for the
case of coupled dark energy simulations normalised to the same
perturbations amplitude.

Nonetheless, we also found that the differences in the CV properties
among these different models significantly change when the CV are
identified in the distribution of collapsed halos rather than in the
CDM distribution itself.  These results suggest that, contrary to what
has been claimed in some other recent works \citep[see
e.g. ][]{sutter2014cDE}, the bias of the tracers of the density field
employed to identify the CV might have a significant impact on the
possibility of using the obtained CV catalogs to probe
cosmology. Therefore, in the present work we will show that a random
subsampling of a simulated CDM distribution to match the density of
tracers expected for any given galaxy survey does not actually provide
a faithful representation of the discriminating power of the survey
with respect to different competing cosmological models.

The paper is organised as follows. In Section~\ref{cDE} we briefly
describe the cDE models considered in the present work and we recall
the main features of the {\small CoDECS} runs.  In Section~\ref{meth}
we describe the void finder algorithm and our method of analysis, and
in Section~\ref{VoidCoD} we present the properties of the CV in the
{\small CoDECS} simulations. Our conclusions are summarised in
Section~\ref{Concl}.

\section{Coupled Dark Energy cosmologies}
\label{cDE}

\subsection{The models}

In this work we aim at studying the statistical and structural properties
of CV in the context of coupled dark energy (cDE)
cosmologies.  In these models, dark energy is represented by a
classical scalar field $\phi$ moving in a self-interaction potential
$V(\phi)$ and directly interacting with CDM particles through an
exchange of energy-momentum, quantified by a coupling function
$\beta(\phi)$. Here we will give only a very essential summary of the
main features of cDE models, and we refer the reader to
\citet{Amendola_2000,Baldi_2011a,CoDECS} for a more thorough
discussion.

The background dynamics of cDE cosmologies is described by the set of
equations:
\begin{eqnarray}
 \dot{\rho}_r+4H\rho_r & = & 0\;,\label{eqn:r} \\
 \dot{\rho}_b+3H\rho_b & = & 0\;,\label{eqn:b} \\
 \dot{\rho}_c+3H\rho_c & = & -\sqrt{\frac{2}{3}}\beta_c(\phi)\frac{\rho_c\dot{\phi}}{M_{Pl}}\;,\label{eqn:c}\\
 \ddot{\phi}+3H\dot{\phi}+V^{\prime}(\phi) & = & \sqrt{\frac{2}{3}}\beta_c(\phi)\frac{\rho_c}{M_{Pl}}\;,
 \label{eqn:phi}
\end{eqnarray}
where the subscripts $r$, $b$, $c$ and $\phi$, indicate the energy
densities $\rho$ of radiation, baryons, CDM, and the dark energy field
$\phi$, respectively, and where the Hubble function is given as usual
by
\begin{equation}
 H^2=\frac{8\pi G}{3}\left(\rho_{\rm r}+\rho_{\rm c}+\rho_{\rm b}+\rho_\phi\right)\;,\label{eqn:H}
\end{equation}
with $M^2_{Pl}\equiv 1/8\pi G$ being the reduced Planck mass. In the
above equations the field $\phi$ is expressed in units of $M_{\rm Pl}$
and an overdot represents a derivative with respect to cosmic time
while a prime denotes a derivative with respect to the field itself.
The source terms on the right-hand side of
Eqs.~\ref{eqn:c} and \ref{eqn:phi} define the interaction between the dark
matter and the dark energy components, with a strength given by the
coupling function $\beta_c(\phi)$.

At the level of linear density fluctuations, the interaction modifies
the gravitational instability processes that govern the evolution of
perturbations as a consequence of a long-range {\it fifth} force
mediated by the dark energy field and acting between CDM fluid
elements.  In the Newtonian limit and on sub-horizon scales, these
effects turn into the following set of modified linear equations
\citep{Amendola2004,Pettorino2008,Baldi_2011c}:
\begin{eqnarray}
 \ddot{\delta}_{\rm c} & = & -2H\left[1-\beta_c\frac{\dot{\phi}}{\sqrt{6}H}\right]\dot{\delta}_{\rm c}+4\pi
G[\bar{\rho}_{\rm b}\delta_{\rm b}+\bar{\rho}_{\rm c}\delta_{\rm c}\Gamma_{\rm c}]\;,\\
 \ddot{\delta}_{\rm b} & = & -2H\dot{\delta}_{\rm b}+
 4\pi G[\bar{\rho}_{\rm b}\delta_{\rm b}+\bar{\rho}_{\rm c}\delta_{\rm c}]\,,
\end{eqnarray}
where $\bar{\rho}_i$ represents the background density of the $i$-th fluid
and $\delta_i\equiv\delta\rho_i/\bar{\rho}_i$ its density
perturbation. The factor $\Gamma_c\equiv 1+4\beta^2_c/3$ represents
the additional fifth force appearing only in the CDM equation while
the term $\beta_c\dot{\phi}$ is a velocity-dependent acceleration
arising as a consequence of momentum conservation. Similar additional
terms characterise the interaction among a discrete set of CDM
particles in the non-linear regime \citep[see][]{Baldi_etal_2010}.

\subsection{The {\small CoDECS} simulations}
\label{s}

\begin{table*}
\caption{A summary of the cosmological models investigated in the
  present work and their main parameters. See \citet{CoDECS} for
  details.}
\begin{tabular}{llcccc}
\hline
Model & Potential  &  
$\alpha $ &
$\beta (\phi )$ &
$w_{\phi }(z=0)$ &
$\sigma _{8}(z=0)$\\
\hline
$\Lambda $CDM & $V(\phi ) = A$ & -- & -- & $-1.0$ & $0.809$ \\
EXP003 & $V(\phi ) = Ae^{-\alpha \phi }$  & 0.08 & 0.15 & $-0.992$ & $0.967$\\
EXP008e3 & $V(\phi ) = Ae^{-\alpha \phi }$  & 0.08 & $0.4 \exp [3\phi ]$& $-0.982$ & $0.895$ \\
SUGRA003 & $V(\phi ) = A\phi ^{-\alpha }e^{\phi ^{2}/2}$  & 2.15 & -0.15 & $-0.901$ & $0.806$ \\
\hline
\end{tabular}
\label{tab:models}
\end{table*}

For our investigation we will make use of the publicly available data
of the {\small CoDECS} simulations \citep[][]{CoDECS} that represent
the largest suite of cDE simulations to date.  These simulations are
carried out with a suitably modified version of the TreePM N-body code
{\small GADGET} \citep[][]{gadget-2} that self-consistently implements
all the above mentioned effects characterising cDE cosmologies
\citep[][]{Baldi_etal_2010}.

For the present work, we will employ the outputs of the {\small
  L-CoDECS} simulations, which follow the evolution of $1024^3$ CDM
particles and as many baryonic particles in a periodic cosmological box of 1
comoving Gpc/h a side. Both CDM and baryonic particles are treated as
collisionless particles, but they experience different accelerations
as a consequence of the interaction between the CDM and the dark
energy fields.

The {\small CoDECS} suite includes six different cosmological models,
for of them are considered in this paper: the reference $\Lambda$CDM
cosmology, a cDE model (EXP003) characterised by a constant positive
coupling $\beta_c>0$ and an exponential self-interaction potential of
the form $V(\phi)=A\exp{(-\alpha\phi)}$, a further model (EXP008e3)
with the same potential but with an exponential coupling,
$\beta_c(\phi)=\beta_0\exp{(\beta_1\phi)}$, and a final scenario
(SUGRA003) with a constant negative coupling, $\beta_c<0$ and a SUGRA
\citep{Brax1999} self-interaction potential
$V(\phi)=A\phi^{-\alpha}\exp{(-\phi^2/2)}$.  A summary of the models
parameters is shown in Table~\ref{tab:models}. All the models have the
same amplitude of perturbations at $z=z_{\rm CMB}$, resulting in a
different amplitude of linear density perturbations at the present
epoch (and consequently  different values of $\sigma _{8}$).

In the present work we will also make use of the public halo catalogs
of the {\small CoDECS} simulations, that have been generated through a
Friend-of-Friend (FoF) algorithm with a linking length of $0.2$ times
the mean inter-particle separation.

\section{Methodology}
\label{meth}

We employ the publicly available void finder {\small VIDE}
\citep[][]{VIDE2015} to identify CV in the CDM and halo distributions
extracted from the snapshots of the {\small CoDECS} simulations within
the different cosmological models described above.  {\small VIDE}
embeds the {\small ZOBOV} algorithm, which allows to identify
depressions in the density distribution of a set of points.  In the
following, we provide a very short summary of how {\small ZOBOV}
works, and we refer to the original {\small ZOBOV} paper
\citep{neyrinck2008} for a more detailed discussion.

Firstly, {\small ZOBOV} associates a cell to each tracer (a
CDM particle or a halo) using a Voronoi tessellation scheme, i.e. the cell $c$
associated to the particle (or halo) $p$ is defined as the region of the box
which is closer to $p$ than to any other particle (or halo) in the box.
Secondly, the algorithm identifies local density minima among these
cells: a density minimum is defined as a Voronoi cell with a lower
density (i.e. a larger volume) than all other cells around it.
Thirdly, {\small ZOBOV} joins together the Voronoi cells surrounding a
local density minimum until cells with larger and larger density are
found, and it identifies CV as the union of these cells.  CV are
joined together via the Watershed Transform \citep[see][]{Platen2007},
which naturally creates a hierarchy in the structures of CV.  All
these procedures are performed also by the {\small ZOBOV} version
included in the {\small VIDE} toolkit. Additionally, {\small VIDE} provides several
different void catalogs for which various types of sample selections
(as e.g. different cuts on the void density contrast or on the void central overdensity)
are applied on top of the original {\small ZOBOV} sample. 
In particular, as CV are found to define a complex hierarchy, with
smaller voids being embedded in larger ones, {\small VIDE} provides
for each identified void the corresponding hierarchy level, and
according to this classification a sample of {\em main} CV
(i.e. those CV that are not embedded in larger voids and that
represent the top of their own void hierarchy) is produced.  {We employed a slightly modified version of
this selection procedure (see Baldi et al. in prep.) to remove pathological CV from the catalog and obtain a more statistically robust and convergent
sample of main CV}.

Finally, since local density minima can also be found in overdense
regions, we decide to remove from the main void catalogs the CV
with a density minimum larger than $20\%$ of the mean density of
the Universe (which is one of the standard cuts provided by {\small
  VIDE}), in order to select only well defined CV for our
comparisons.  For each identified void, {\small ZOBOV} also calculates
the probability that the void might arise in a uniform Poissonian
distribution of points, which is directly related to the density
contrast between the minimum density of the void and its boundary.  As
this density contrast is provided for each void also by the {\small
  VIDE} catalog, we remove CV with a density contrast below 1.57,
corresponding to a probability of arising as Poisson noise
larger than 2$\sigma$ \citep[see][]{neyrinck2008}.\\

{\small VIDE} defines CV as spherical regions centered in the
barycenter, $\vec{x}_c$, of the underdense regions provided by {\small
  ZOBOV}, where:
\begin{equation}
\label{center}
%\vec{x}_c = \frac{1}{V_{\rm VOID}} \sum\limits_{i=1}^N \vec{x}^{p}_i
%\cdot V^{p}_i \,,
\vec{x}_c = \frac {\sum\limits_{i=1}^N \vec{x}^{p}_i \cdot V^{p}_i }
{\sum\limits_{i=1}^N V^{p}_i } \,,
\end{equation}
and $\vec{x}^{p}_i$ and $V_i$ are the positions of the $i-th$ tracer
and the volume of its associated Voronoi cell, while $N$ is the number
of tracers included in the void.  The radius of the sphere (i.e. the
effective radius of the void, $R_{\rm eff}$) is then computed from the
overall volume of the underdense region by assuming sphericity:
\begin{equation}
V_{\rm VOID} \equiv {\sum\limits_{i=1}^N V^{p}_i }= \frac{4}{3} \pi R_{\rm eff}^3 \,.
\end{equation}

It has been shown that different void finders based on dynamical
criteria, instead of density or geometry criteria, might reduce the
shot noise error \citep{elyiv2015}. Nevertheless, the CV finder used
here is accurate enough for the purpose of the present analysis, as we
investigate the main properties of large CV extracted from dense
numerical simulations.

\section{The statistics of voids in the CoDECS}
\label{VoidCoD}

With the catalogs of CV extracted from the {\small CoDECS} simulations
as described in the previous Section~at hand, we perform some basic
analyses of the statistical and structural properties of the CV in the
different cosmologies, namely the CV filling factor (i.e. the fraction
of the cosmic volume occupied by CV), their size distribution
(i.e. the abundance of CV as a function of their size), and their
stacked radial density profiles, and compare these observables to the
reference $\Lambda $CDM case \citep[see also ][]{Li_2011}. We perform
such comparison for CV identified both in a randomly subsampled CDM
density field and in the distribution of collapsed halos, to highlight
how the use of tracers with different bias might result in a different
relative behaviour of the models.

\subsection{Void statistics in the CDM distribution}
\label{CDM-CoDECS}

\begin{figure}
\begin{center}
\includegraphics[scale=0.5]{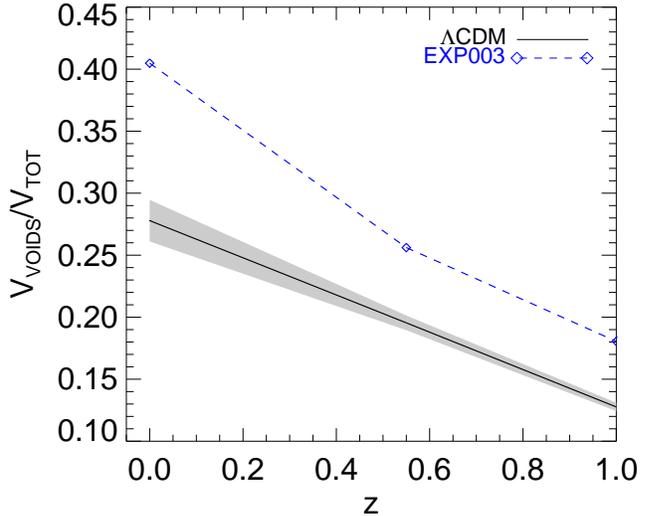}
\caption{The redshift evolution of the volume fraction of CV
  identified in the CDM distribution for the $\Lambda$CDM (black
  solid line) and EXP003 (blue dashed line) models. The shaded area
  (shown only for $\Lambda$CDM) represents the uncertainty, computed
  with the jackknife method.}
\label{DM_vol_frac}
\end{center}
\end{figure}

Let us start by considering the CV catalogs extracted from the CDM
density field, i.e. directly from the {\small CoDECS} snapshots at
different redshifts. To better handle the simulation data we have
made use of the subsampling routine included in {\small VIDE} to
randomly subsample the CDM particles of the simulation snapshots down
to an average density of $2 \times 10^{7}$ particles per cubic \Gpch.
For this comparison we will focus only on two out of the four models,
namely the reference $\Lambda $CDM cosmology and the EXP003 scenario, 
which is the most extreme realisation (in
terms of deviations at the background and linear perturbations level)
of cDE models that we have at our disposal.

First of all, we compare the evolution of the volume fraction occupied
by CV at different cosmic times, also known as the void filling
factor, to check whether the interaction between DE and CDM particles
implemented in our extreme cDE model has an impact on such fraction.
Fig.~\ref{DM_vol_frac} displays the evolution of the CV volume
fraction, $V_{\rm VOIDS}/V_{\rm TOT}$, where $V_{\rm VOIDS}$ is the
sum of all the main CV volumes in a given snapshot of the simulation
and $V_{\rm TOT}$ is the total volume of the box, i.e. $1$
$h^{-3}\,\mbox{Gpc}^3$. The statistical error, shown in
Fig.~\ref{DM_vol_frac} by the shaded grey region around the
$\Lambda$CDM line, has been computed with a jackknife method.

As expected, the volume fraction of CV increases with time,
irrespectively of the underlying cosmological model, due to
gravitational instability.  Moreover, as one can see from
Fig.~\ref{DM_vol_frac}, the volume fraction occupied by CV in the cDE
model EXP003 is significantly larger than in the reference $\Lambda
$CDM cosmology, reflecting the higher normalisation of the amplitude
of linear perturbations at low redshifts in EXP003.  More
quantitatively, the volume fraction in EXP003 is roughly $40\%$ larger
than the corresponding $\Lambda $CDM fraction, at all redshifts
between $z=1$ and $z=0$. Clearly, the observed differences between the
cDE model and the standard $\Lambda $CDM cosmology are statistically
significant.

\begin{figure*}
\includegraphics[width=\textwidth]{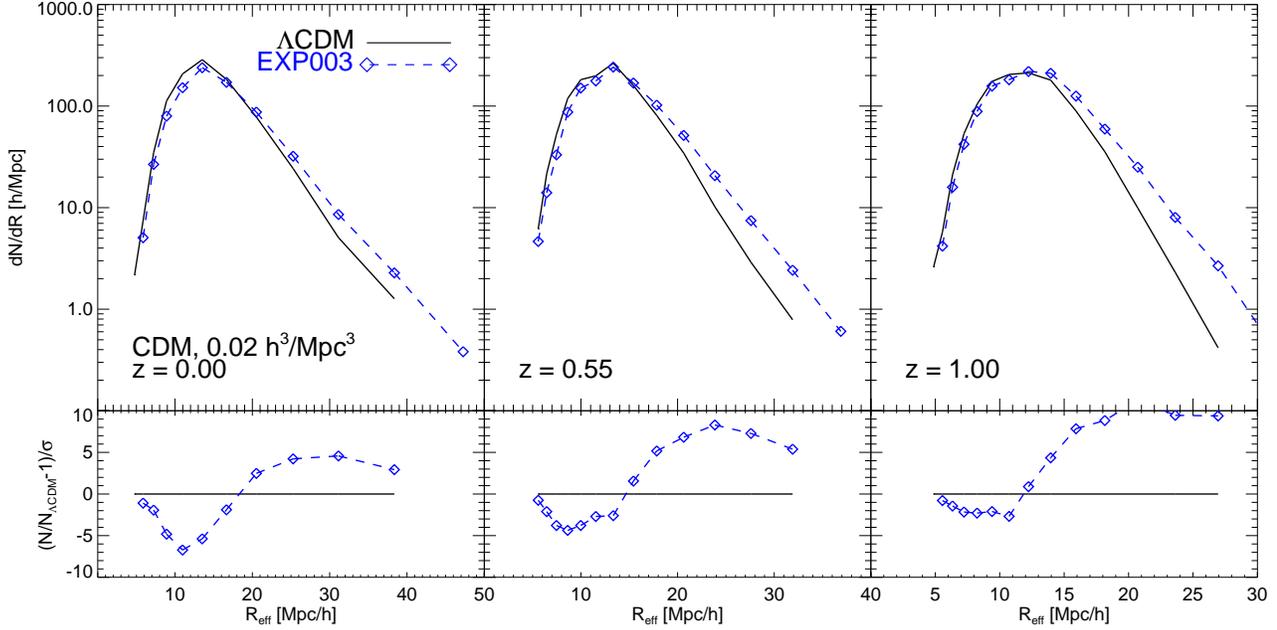}
\caption{Top panels: the size distribution of CV in the CDM
  distribution for the $\Lambda$CDM (black solid line) and EXP003
  (blue dashed lines) models. Bottom panels: the relative
  differences between the two models in units of the standard
  deviation $\sigma$, computed for the $\Lambda$CDM model.}
\label{fig:CDM_VSD_CMB}
\end{figure*}

As a second step, we compare the relative abundance of CV as a
function of their size, by computing in the two cosmological models
the differential size distribution (hereafter DSD), defined as the
number of CV with an effective radius $R_{\rm eff}$ falling within a
set of size bins. In the upper panels of Fig.~\ref{fig:CDM_VSD_CMB} we
show the DSD at three different redshifts ($z=\left\{ 0\,, 0.55\,,
  1\right\} $, from left to right) for the two cosmological models
(black solid lines for $\Lambda $CDM and blue dashed lines for
EXP003), while in the bottom panels we show the relative difference
with respect to the reference $\Lambda $CDM cosmology (in units of its
statistical error $\sigma$). As one can see in the figure, at $z=0$
the number of CV in the cDE cosmology with $R_{\rm eff} \gtrsim 20$
\Mpch is at least $50\%$ larger than in $\Lambda$CDM: this difference
corresponds to more than $4 \sigma$.  At $z=0.55$ the same ratio
applies to CV with $R_{\rm eff} \gtrsim 15$ \Mpch, and at $z=1$ to
CV with $R_{\rm eff} \gtrsim 12$ \Mpch, with differences
corresponding to $5 \sigma$ and $7 \sigma$, respectively.  Therefore,
also in this different statistics the two models are clearly
distinguishable from each other.

As a third statistic of our CV samples, we investigate the average
stacked density profiles of CV having a comparable size. Many
recent works \citep[e.g.][]{ricciardelli2013, ricciardelli2014,
  Hamaus2014} suggested that the average profile of CV is
self-similar in the standard $\Lambda$CDM cosmology, which makes CV
an ideal target for geometrical tests such as the AP test. Therefore,
we now aim at investigating whether the interaction between DE and CDM
particles might induce some additional features on the density profile
of CV.

\begin{figure*}
{\includegraphics[width=0.48\textwidth]
  {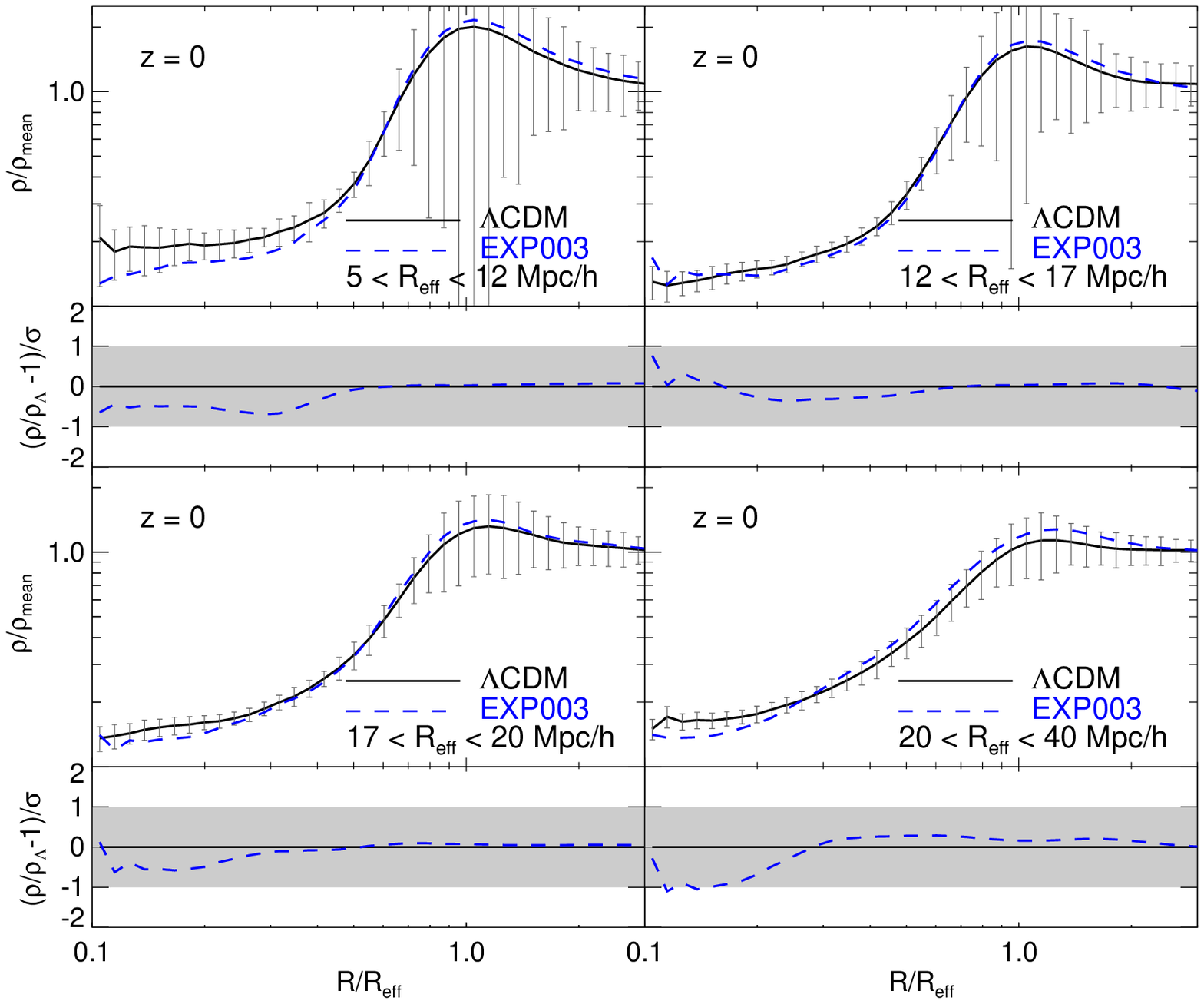}}
  {\includegraphics[width=0.48\textwidth]{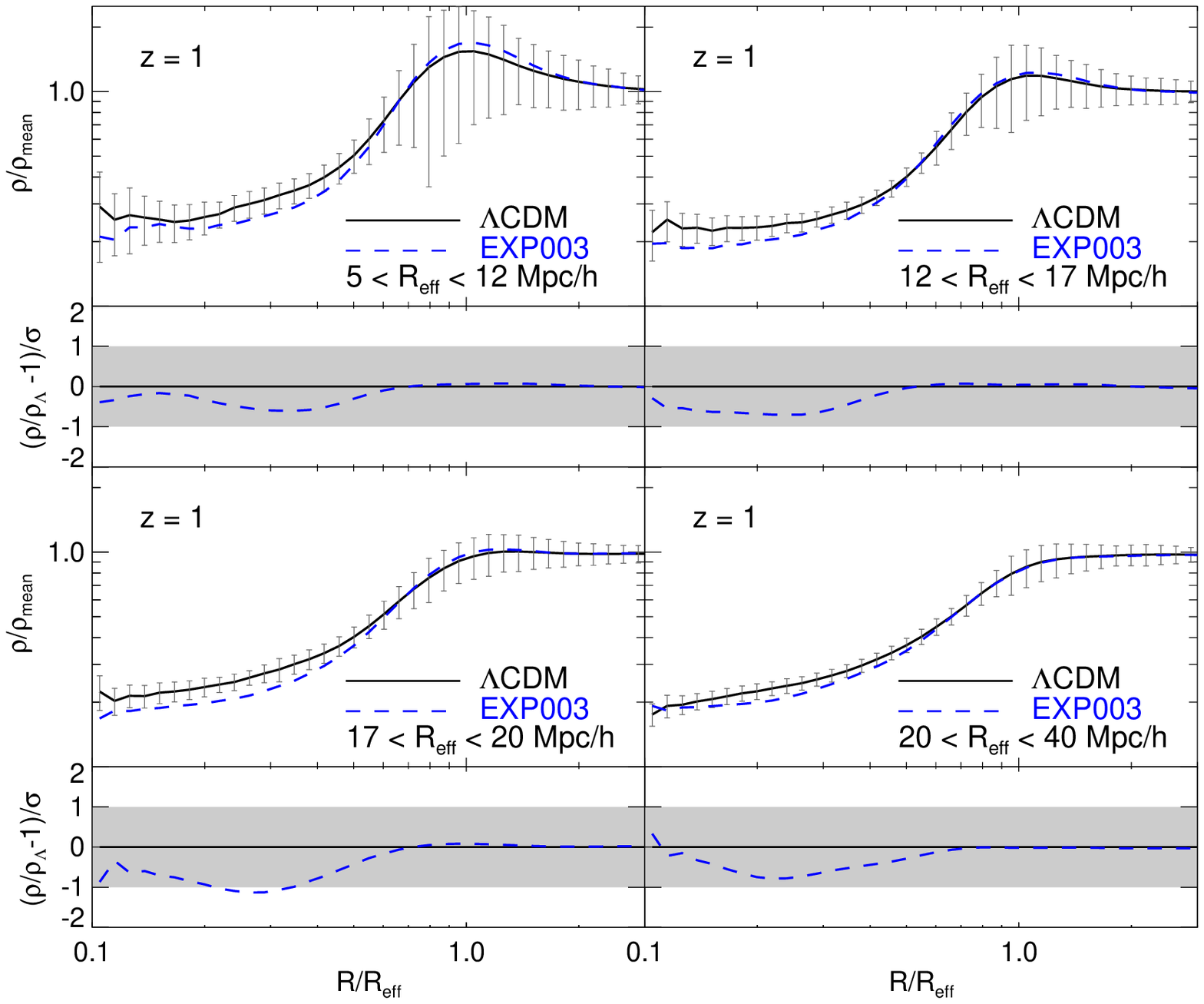}}
  	\caption{
The stacked profiles of CV in the CDM distribution
          for the $\Lambda$CDM (black solid lines) and EXP003 (blue
          dashed lines) models. Results are displayed at two different
          redshifts, $z=0$ and  $z=1$ (left and right blocks of
          panels, respectively) for four ranges of $R_{\rm eff}$, as
          labeled. {The error bars indicate the corrected sample standard deviation in each radial bin computed on the 100 randomly selected CV, while
          the sub-panels display the relative difference between the profiles in units of the statistical significance of the averaged profile.}
}
        \label{fig:CDM_prof}
\end{figure*}

To this end, we first compute the spherically-averaged radial density
profile of each individual void by estimating the CDM density within a
series of logarithmically equispaced spherical shells centred in the
barycentre of each void and normalised to the void effective radius
$R_{\rm eff}$.  The profiles are then stacked for CV with similar
$R_{\rm eff}$.  Since the profiles of each void is calculated in units
of $R_{\rm eff}$ in the first place, the stacking procedure basically
consists in the calculation of the mean profile in each logarithmic
radial bin.  We randomly included $100$ CV for each bin.  The
results are presented in Fig.~\ref{fig:CDM_prof}, where we show the
comparison of the stacked density profiles obtained using the CDM void
catalog in the standard $\Lambda$CDM cosmology and in the EXP003
model. {In the upper panels the error bars represent the corrected 
sample standard deviation computed on the 100 randomly selected CV.}
The shape of the profiles is qualitatively the same as found
in previous works \citep[][]{ricciardelli2013, ricciardelli2014,
  Hamaus2014}: we observe a deep underdensity at $R \rightarrow 0$ and
a compensative overdensity at $R \rightarrow R_{\rm eff}$. At $R > 1.5
\cdot R_{\rm eff}$ the profiles reach the mean density of the
Universe.
{In the lower panels, we plot the relative difference between the models in units
of the statistical significance $\sigma $ computed as the sample standard deviation propagated to the relative difference. 
The grey shaded area represents a $\pm 1\sigma $ significance.}

  As the figure clearly shows, the stacked profiles of EXP003 do not
  show significative differences from $\Lambda$CDM at the considered
  redshifts, with deviations always lying well within the mean square
  error. Nonetheless, we can observe that, at {\bf $R \rightarrow 0$},
  EXP003 generally shows a density $10-25 \%$ smaller than
  $\Lambda$CDM, thereby showing that CV are {\em emptier} in cDE.
  On the other hand, the compensative over-density at $R=R_{\rm eff}$
  for EXP003 looks generally more prominent than $\Lambda$CDM (except
  for CV with $5 < R_{\rm eff}[$\Mpch$]< 8$). Therefore,
    although with a low statistical significance, the central regions
    of the main CV appear to be more underdense in cDE models than in
    $\Lambda $CDM, which is expected to result in a corresponding
    stronger signal in void lensing surveys.

\subsection{Void statistics in the halo distribution} 
\label{voidChalo}
We will now repeat the same three statistics of the CV properties
discussed in Section~\ref{CDM-CoDECS} for the CV catalogs obtained by
running {\small VIDE} on the distribution of FoF halos extracted from
the {\small CoDECS} simulations at the same three redshifts
investigated before (i.e. $z=\left\{ 0\,, 0.55\,, 1\right\} $). The
use of the FoF halos as tracers of the matter distribution has the
appealing property to mimic real observations, where CV are
identified in the distribution of luminous galaxies. {In particular, we have made use
of the publicly available {\small CoDECS} halo catalogs  that have been
obtained through a FoF algorithm with a linking length $0.2$ times the mean inter-particle separation.}
As we will show
below, the differences between the cDE model EXP003 and the standard
$\Lambda $CDM cosmology in all the three statistics are much weaker
than what previously found for the CDM distribution. For this reason,
we will include in this comparison also other two cDE models available
within the {\small CoDECS} suite, namely the EXP008e3 and the SUGRA003
models (introduced in Section~\ref{s}), in order to verify whether
different realisations of the cDE scenario might have a stronger
impact on the CV defined by the FoF halo distribution than the EXP003
model. Our comparison will show that this is actually not the case, as
expected from the fact that EXP003 is the most extreme of the {\small
  CoDECS} models in terms of background and linear deviations from
$\Lambda $CDM.

\begin{figure}
\begin{center}
\includegraphics[scale=0.5]{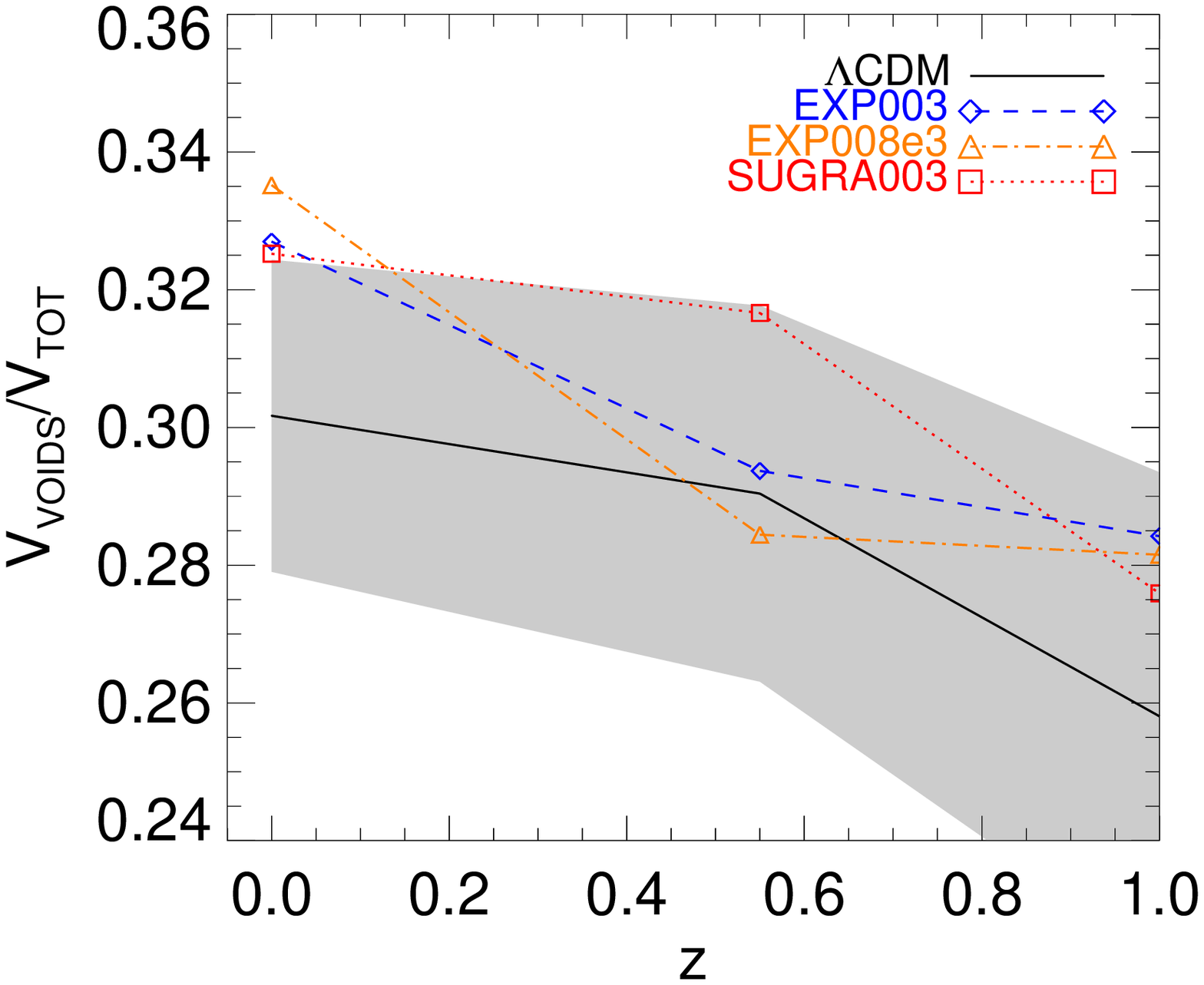}
\caption{ The redshift evolution of the volume fraction of CV
  identified in the halo distribution for different cosmological
  models: $\Lambda$CDM (black solid line), EXP003 (blue dashed line),
  EXP008e3 (dot-dashed orange line), SUGRA003 (red dotted line).  The
  shaded area (shown only for $\Lambda$CDM) represents the
  uncertainty, computed with the jackknife method.}
\label{vol_frac}
\end{center}
\end{figure}

First of all, in Fig.~\ref{vol_frac} we compare the CV filling factor
for these new void catalogs, as already done in Fig.~\ref{DM_vol_frac}
for the CV in the CDM distribution. The void volume and the
dispersion indicated by the grey shaded area are computed as outlined
above.  The figure shows, as expected, that the void volume fraction
increases in time, and that the $\Lambda$CDM model has generally the
lowest volume fraction with respect to the other cDE models that are
characterised by a higher normalisation of the linear power spectrum.
Nonetheless, as the figure clearly shows, these differences are now
much smaller and lie within the 3$\sigma$ statistical dispersion so
that no significant differences in the CV filling factor appear among
the various cDE models and the standard $\Lambda $CDM cosmology at all
redshifts.
This result is starkly different from what found for the CV in the CDM
distribution for the EXP003 model.

\begin{figure*}
\includegraphics[width=\textwidth]{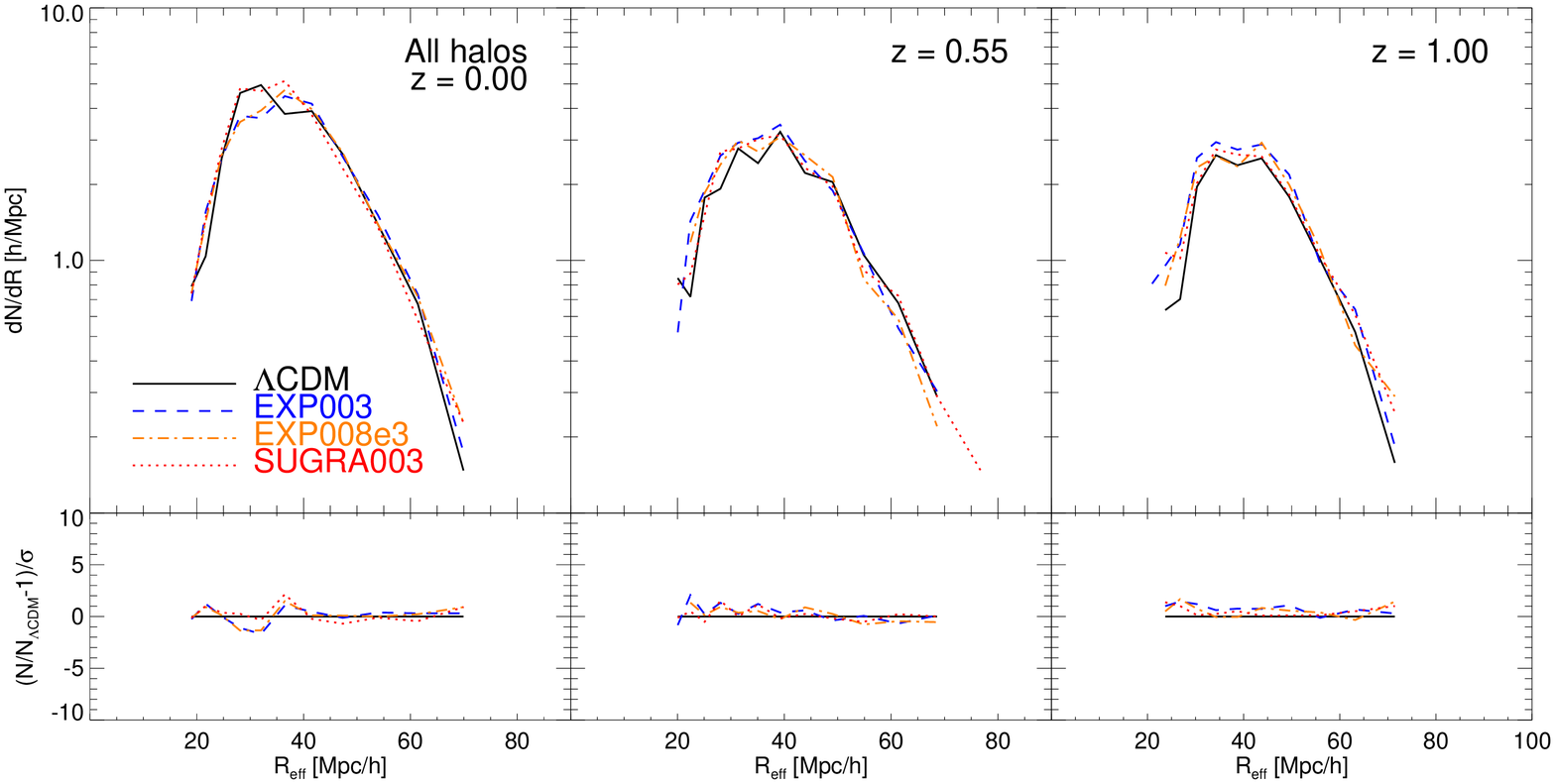}
\caption{Top panels: the size distribution of CV in the halo
  distribution for the $\Lambda$CDM (black solid line), EXP003 (blue
  dashed lines), EXP008e3 (orange dot-dashed lines) and SUGRA003 (red
  dotted lines) models. Bottom panels: the
  relative differences between the cDE models and the $\Lambda$CDM
  one, in units of the standard deviation $\sigma$, computed for the
  $\Lambda$CDM model.}
\label{fig:halos_VSD}
\end{figure*}

In Fig.~\ref{fig:halos_VSD} we then display the DSD for CV identified
in the distribution of FoF halos, analogously to what done in
Fig.~\ref{fig:CDM_VSD_CMB} for the CDM distribution.  The DSD is shown
for the different models ($\Lambda$CDM by black solid line, EXP$003$
by blue dashed line, EXP008e3 by orange dot-dashed line and SUGRA003
by red dotted line) in the upper panels, while the bottom panels
report the percent deviation in units of the statistical significance
$\sigma$ from the reference $\Lambda$CDM case.  While at $z=0$ and
$z=0.55$ no significant differences appear among the models, one can
observe an excess of small CV for the EXP003 model at $z=1$. At
this redshifts EXP003 shows $\sim 50 \%$ more CV with $R_{\rm eff}
< 30$ \Mpch than $\Lambda$CDM, although within a confidence of $1
\sigma$.

This is again a very different result with respect to what
previously found for the CV identified in the CDM distribution,
where the largest differences with respect to the standard
cosmological model appeared at the large size tail of the
distribution.  It should however be noticed that due to the different
density of the tracers between the subsampled CDM distribution adopted
in the previous Section~and the FoF halo distribution shown here, the
mean separation between particles and hence the average size of CV is
different in the two cases. Therefore, the size range that appeared as
the large-size tail for the CDM CV ($20 < R_{\rm eff} [$\Mpch$]<
30$) is now representing the small-size part of the void samples of
the FoF halo distribution.  For this reason, the two results might
still appear consistent with each other despite their different
qualitative trends.  To address this issue, in Section~\ref{bias}
below we will compare the DSD of CV identified in a different
random subsample of the CDM particles distribution with the same
density of tracers as the FoF halo catalogue.  Nonetheless, the clear
differences between the background evolution of $\Lambda$CDM and cDE
models (see Fig.~\ref{fig:CDM_VSD_CMB}) are not expected to be
detected and do not appear for CV in halos.

\begin{figure*}
{\includegraphics[width=0.48\textwidth]
      {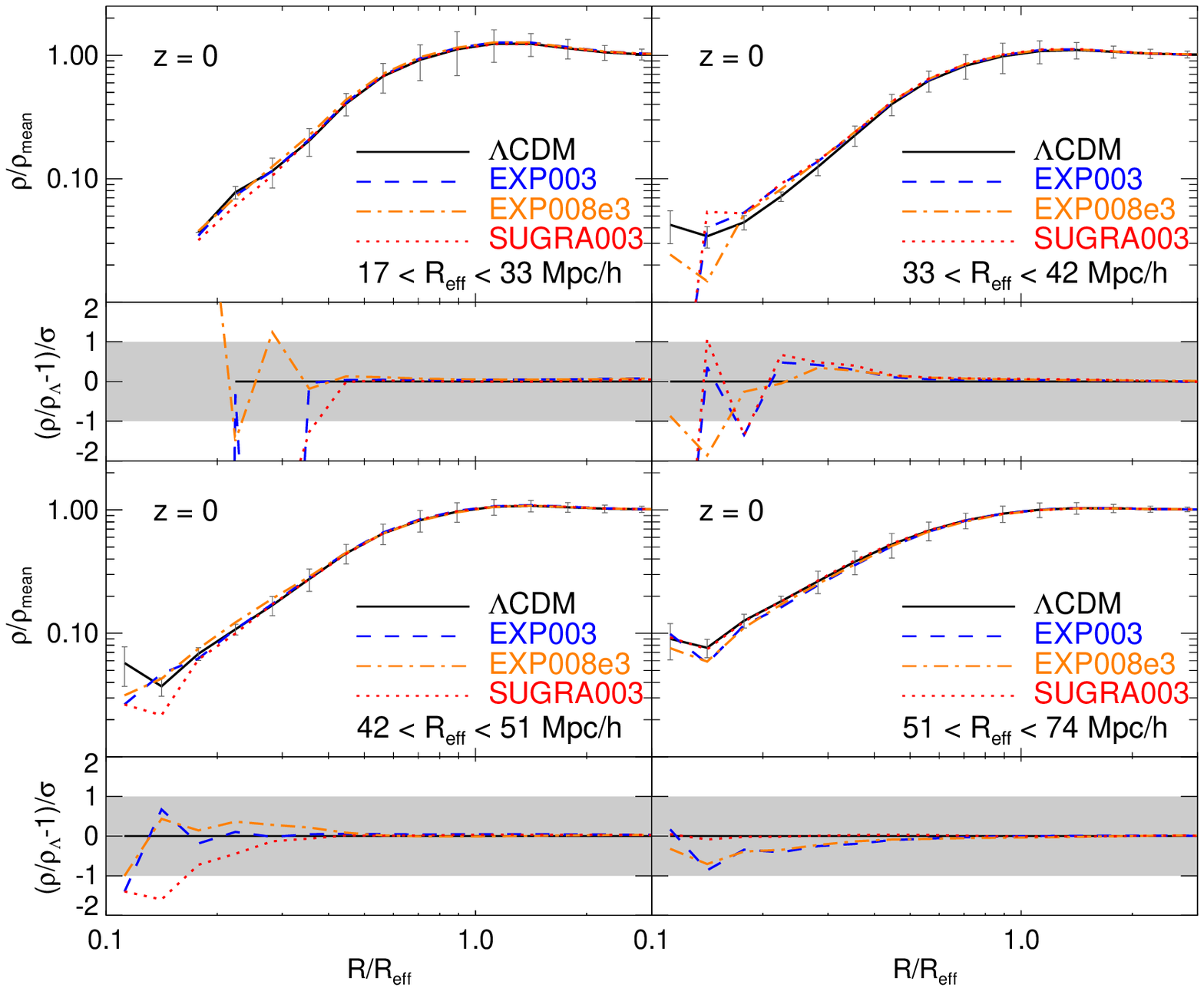}}
{\includegraphics[width=0.48\textwidth]{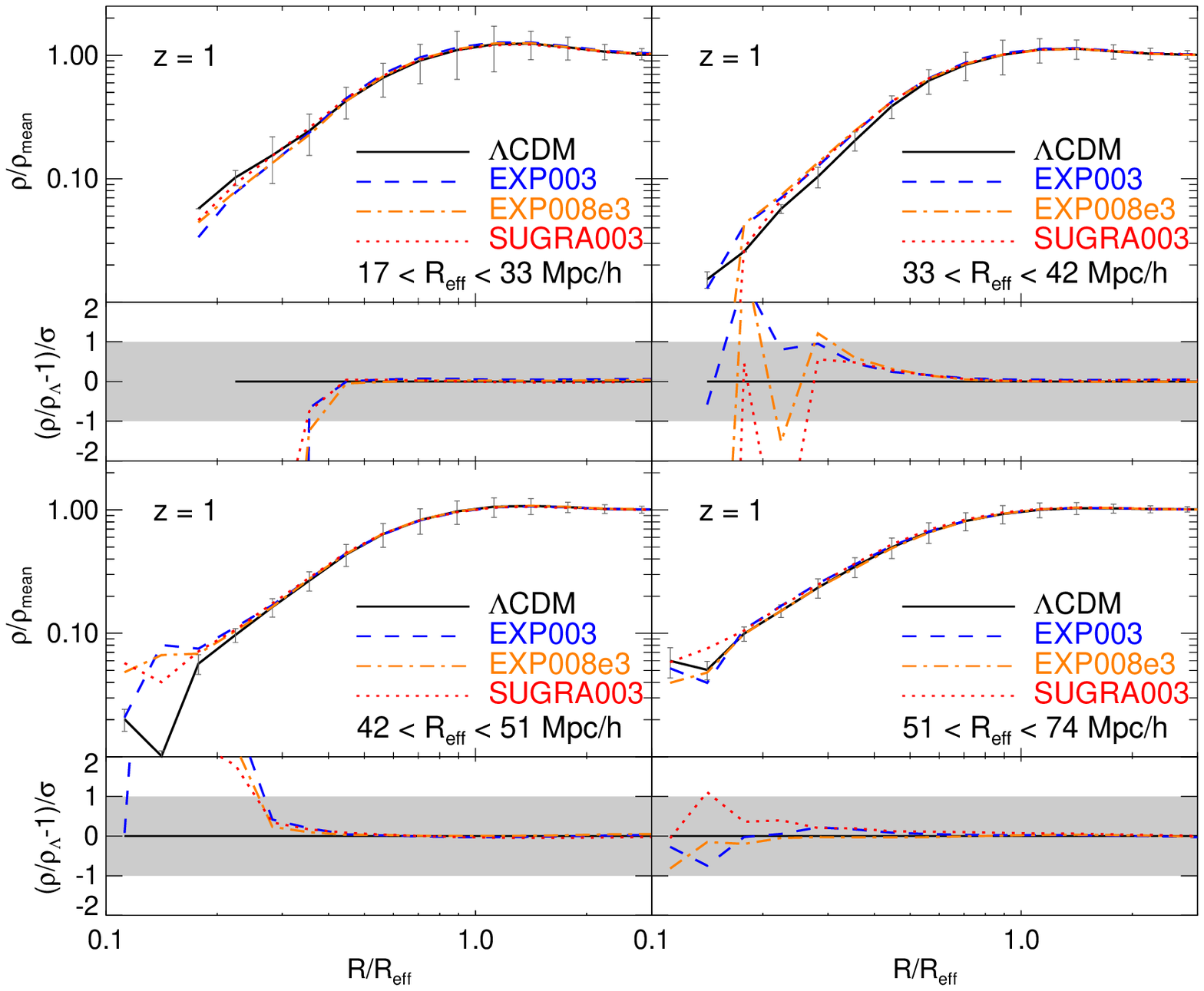}}
\caption{
The stacked profiles of CV in the halo distribution
          for the $\Lambda$CDM (black solid lines), EXP003 (blue
          dashed lines), EXP008e3 (orange dot-dashed lines) and
          SUGRA003 (red dotted lines) models. Results are displayed at two different
          redshifts, $z=0$ and  $z=1$ (left and right blocks of
          panels, respectively) for four ranges of $R_{\rm eff}$, as
          labeled. {The error bars in the upper panels are computed as for Fig.~\ref{fig:CDM_prof}, and the
          sub-panels display again the relative difference of the profiles with respect to the $\Lambda $CDM one in units of
          the statistical significance of the averaged profile.}
.}
\label{fig:halosprof}
\end{figure*}

Before moving to this additional comparison, we conclude our analysis
of the CV extracted from the FoF halo distribution by comparing the
void stacked density profiles as we did in Fig.~\ref{fig:CDM_prof} for
the CV in the CDM distribution.  In Fig.~\ref{fig:halosprof} we
show the analogous to Fig.~\ref{fig:CDM_prof} for these new CV samples
at $z=0$ and $z=1$.
For both redshifts we do observe significative
deviations from the $\Lambda$CDM profile only in the inner part of the CV, where
shot noise as well as an offset of the computed centre of the void from the real centre
can strongly affect the profiles, as pointed out by e.g. \citet[]{Nadathur2014,
  Nadathur2015}.
We also observe that the over-compensative region
  around $\sim 1 R_{\rm eff}$ is not as prominent as in the CDM CV
  (see Fig.~\ref{fig:CDM_prof}), once again showing differences
  between tracers of density.  We notice that the density minimum does
  not lie exactly at $R \sim 0$, which was not the case for CV in
  CDM.  These last problem might be caused by the definition of centre
  as in eq. \ref{center}, as pointed out by \cite{Nadathur2014,
    Nadathur2015}. Eq. \ref{center}, anyway, is sufficient for our
  purpose because, by using it, we can easily compare our results with
  other works, and because we are not discussing features related to
  the correlation function of CV (in which the position of the
  centre of voids plays a fundamental role).

\subsection{The impact of halo bias}
\label{bias}

\begin{figure*}
\includegraphics[width=\textwidth]{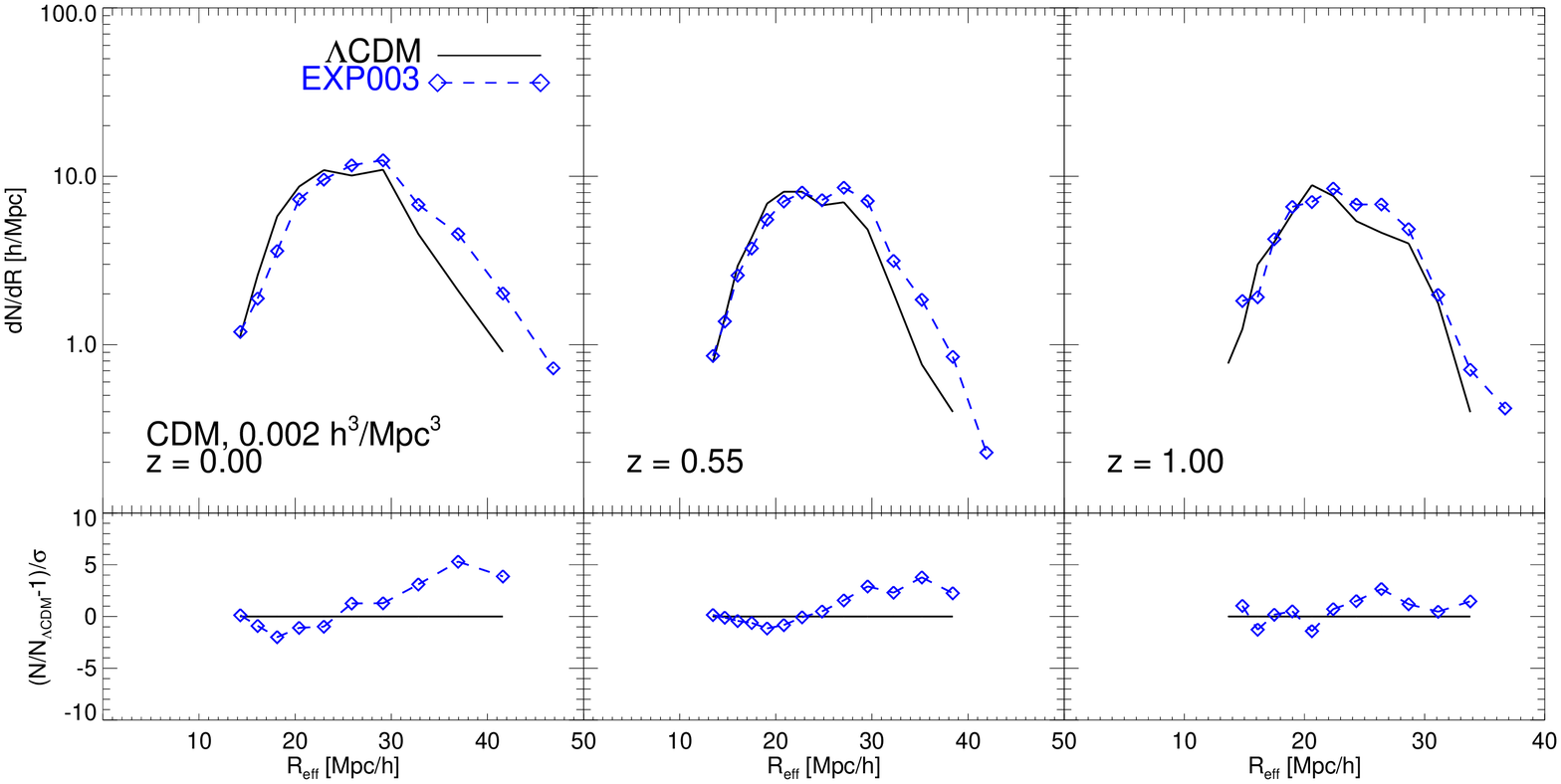}
\caption{Top panels: the size distribution of CV in the CDM
  distribution diluted to the same density of the halo catalogue,
  for the $\Lambda$CDM (black solid lines) and EXP003 (blue dashed
  lines) models. Bottom panels: the relative differences
  between the two models, in units of the standard deviation,
  $\sigma$, computed for the $\Lambda$CDM model.}
\label{fig:CDM_VSD_CMB_halodens}
\end{figure*}

As introduced in Section~\ref{voidChalo}, it is interesting to compare
the abundance of CV as a function of their effective radius, $R_{\rm
  eff}$, for CV samples extracted from the FoF halo catalogs and from
a random subsampling of the CDM distribution having the same number of
tracers in the simulation box as the number of FoF halos. This will
ensure that the mean inter-particle separation of the two samples of
tracers is the same such that the corresponding average size of CV
will be comparable in the two cases, thereby allowing for a direct
comparison of the statistical properties of the two CV catalogs over a
similar range of void sizes.  This approach has been followed in
several recent works \citep[see e.g.][]{Hamaus2014, Sutter_etal_2015,
  Pisani_etal_2015}, where the discriminating power of CV in future
galaxy surveys has been inferred from the expected properties of CV
identified in a random subsample of a simulated CDM distribution
having the same density of the survey under investigation.  To this
end, we have randomly subsampled the CDM distribution of the {\small
  CoDECS} snapshots at the relevant redshifts to a total number of
$1.5 \cdot 10^{6}$ particles, corresponding to the total number of
objects in the FoF halo catalog of the $\Lambda $CDM simulation at
$z=0$.

In Fig.~\ref{fig:CDM_VSD_CMB_halodens} we show the equivalent to
Fig.~\ref{fig:CDM_VSD_CMB} for this new random subsampling and compare
the abundance of CV in the $\Lambda$CDM and EXP003 models. We
observe that, although the range of void sizes is now comparable to
what was shown in Fig.~\ref{fig:halos_VSD}, the comparison between
$\Lambda$CDM and EXP003 still appears starkly different in the two
cases.  Also in this case, as already shown for a denser sample of CDM
tracers, the EXP003 scenario includes a larger number of CV of
large sizes at all redshifts with respect to the standard $\Lambda
$CDM cosmology, with a qualitatively different trend with respect to
what shown in Fig.~\ref{fig:halos_VSD}.  The comparison of
Figs. \ref{fig:halos_VSD} and \ref{fig:CDM_VSD_CMB_halodens} clearly
indicates that CV in the CDM distribution and CV in the
distribution of halos are characterised by different statistical
properties. As the density of the two tracers is the same, these
different properties must be associated with the different bias of the
two samples with respect to the underlying true density field: while a
random subsampling of the CDM distribution is an unbiased tracer of
the density field, halos are biased and the bias is expected to evolve
differently in cDE models than in $\Lambda $CDM
\citep[][]{Marulli_Baldi_Moscardini_2012,Moresco_etal_2014}.
{More quantitatively, the lower bias of the EXP003 model compensates for the 
higher value of the perturbations amplitude. In Table~\ref{tab:bias} we display the 
value of the bias \citep[as computed in][]{Marulli_Baldi_Moscardini_2012} and of $\sigma _{8}$
at various redshifts for the two models. As one can see from the last column, the combination
$b(z)\cdot \sigma _{8}(z)$ is substantially closer between the two models compared to the 
value of $\sigma _{8}$ alone.}
This result
suggests that the assumption (implicitly adopted in many recent works)
that the properties of CV in a subsampled set of CDM particles
extracted from a cosmological simulation can faithfully reproduce the
statistics of CV identified in a galaxy survey is not valid.
\begin{center}
\begin{table}
\begin{tabular}{|c|ccc|ccc|}
\hline
 & & $\Lambda $CDM & & & EXP003 & \\
\hline
\hline
$z$ & $b$ & $\sigma _{8}$ & $b\cdot \sigma_{8}$ & $b$ & $\sigma _{8}$ & $b\cdot \sigma_{8}$\\
\hline
0.00 & 1.2 & 0.809 & 0.971 & 1.046 & 0.967 & 1.011 \\
0.55 & 1.584 & 0.618 & 0.979 & 1.310 & 0.733 & 0.960 \\
1.00 & 2.049 & 0.504 & 1.033 & 1.633 & 0.595 & 0.972 \\
1.60 & 2.903 & 0.398 & 1.155 & 2.235  & 0.468 & 1.046 \\
2.00 & 3.630 & 0.348 & 1.263 & 2.739 & 0.408 &  1.118 \\
\hline
\hline
\end{tabular}
\caption{{The bias $b(z)$ and the  normalisation of the linear perturbations amplitude $\sigma _{8}(z)$ for the $\Lambda $CDM and EXP003 cosmologies. The rightmost column for each model displays the combination $b(z)\cdot \sigma _{8}(z)$, showing how this combination is much similar for the two models as compared to $\sigma _{8}$ alone. As a consequence, the differences in the void populations extracted from the biased tracers within the two scenarios are significantly suppressed with respect to the case of the CV in the CDM distribution.}}
\label{tab:bias}
\end{table}
\end{center}

\begin{figure*}
\includegraphics[width=\textwidth]{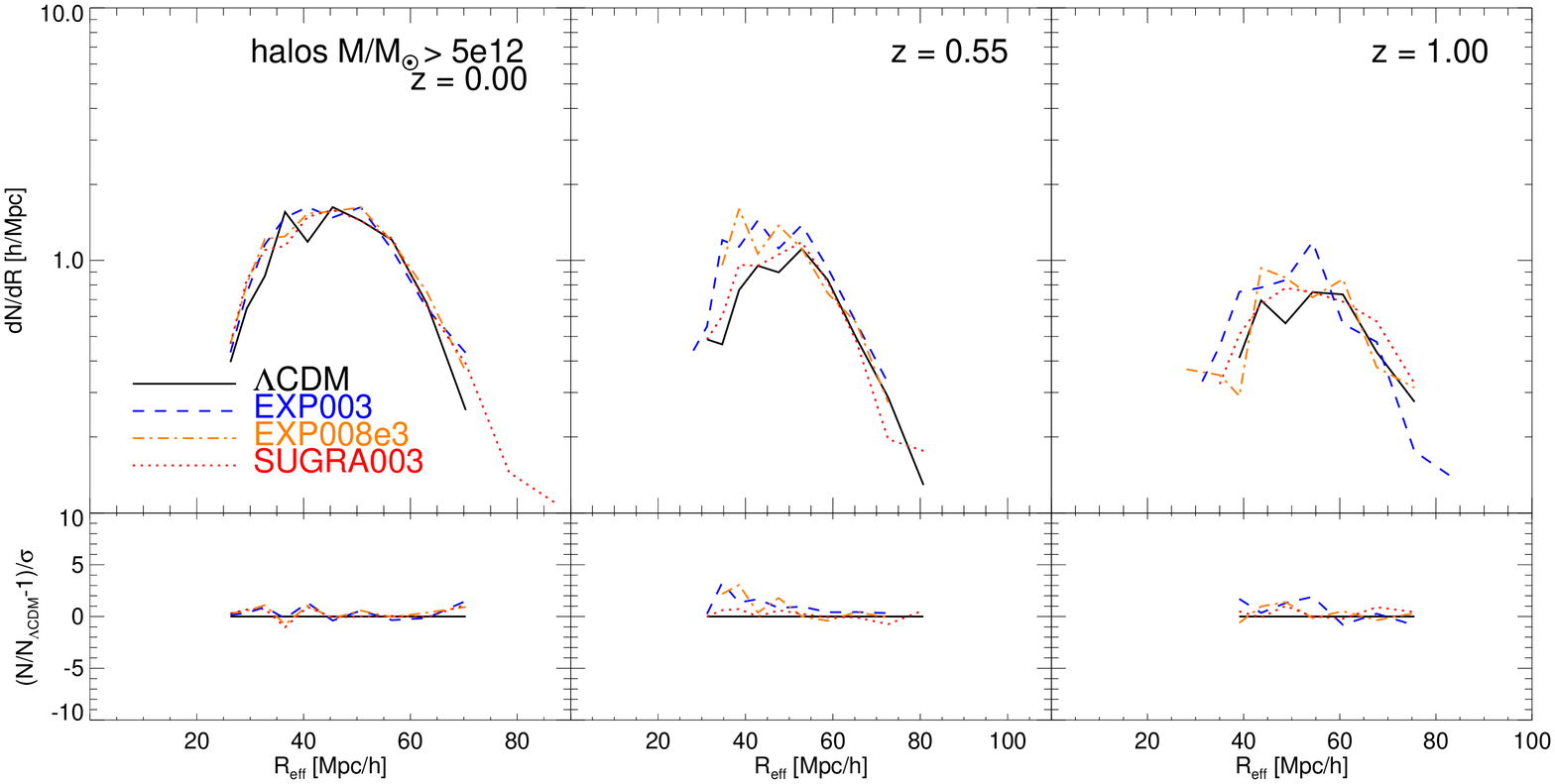}
\caption{Top panels: the size distribution of CV identified in
  the distribution of halos with mass $M > 5 \cdot 10^{12}
  M_{\bigodot}$, for the $\Lambda$CDM (black solid lines), EXP003
  (blue dashed lines), EXP008e3 (orange dot-dashed lines) and SUGRA003
  (red dotted lines) models.  Bottom panels: the relative
  differences between the cDE models and the $\Lambda$CDM one, in
  units of the standard deviation $\sigma$, computed for the
  $\Lambda$CDM model.}
\label{fig:bias_VSD}
\end{figure*}

In order to further validate this result, we compute the DSD of CV
identified in the distribution of FoF halos with masses $M > 5 \cdot
10^{12} M_{\bigodot}$, thus considering tracers with larger masses
and, therefore, with higher bias. This comparison is shown in the
upper panels of Fig.~\ref{fig:bias_VSD}, while the bottom panels
display the deviation (in units of $\sigma$) between the models.  At
$z=0$ we do not observe any significative difference between models
(in agreement with Fig.~\ref{fig:halos_VSD}), while at larger
redshifts we find that the EXP003 cDE model features a larger number
of small CV ($30 < R_{\rm eff} $[\Mpch]$ < 60$) as compared to
$\Lambda$CDM, though the effect is small. The comparison between
Figs.~\ref{fig:CDM_VSD_CMB} and ~\ref{fig:bias_VSD} indicates again
that CV in the CDM distribution and in the distribution of halos
are characterised by different statistical properties.  This result
clearly shows that the bias of the tracers from which CV are
identified has a non-trivial impact on the relative statistical
properties of the CV sample between two competing cosmological
scenarios.  Therefore, when comparing CV in $\Lambda$CDM and cDE
models, CV in halos (which are biased tracers of the underlying
density field) will not provide a faithful representation of how the
models might differ in the properties of CV in the CDM distribution.

\section{Discussion and Conclusions}
\label{Concl}
In this paper we analysed the statistical properties of CV in
$\Lambda$CDM and cDE models. In particular, we compared the properties
of CV detected in the distribution of CDM and in collapsed halos, by
means of a suite of large cosmological simulations, the {\small
  CoDECS}.  We focused on three CV statistics: the filling factor, the
size distribution and the stacked density profiles.

In Section~\ref{CDM-CoDECS} we investigate the properties of CV in the
CDM distribution, considering the $\Lambda$CDM and the cDE model
EXP003, which represents -- among the available {\small CoDECS} models
-- the most extreme case showing the largest discrepancies with
respect to $\Lambda$CDM in several other observables \citep[see
e.g. the results of][]{CoDECS, Lee_Baldi_2011,
  Marulli_Baldi_Moscardini_2012, Beynon_etal_2012,
  Cui_Baldi_Borgani_2012, Giocoli_etal_2012, Carbone_etal_2013,
  Moresco_etal_2014, Pace_etal_2015, Giocoli_etal_2015}.  Our main
results can be summarised as follows.
\begin{enumerate}
\item{The filling factor of CV detected in the CDM distribution in the
    EXP003 model is significantly larger than in the $\Lambda$CDM
    case, as expected due to the higher normalization of the amplitude
    of linear perturbations at low redshift (Fig.~\ref{DM_vol_frac}).
    More quantitatively, the volume fraction in EXP003 is $\sim 40\%$
    larger than the corresponding $\Lambda$CDM fraction: based on a
    jackknife approach, this is detectable with a very high
    statistical significance.}

\item {For what concerns the DSD (Fig.~\ref{fig:CDM_VSD_CMB}), we
    found an excess of large CV in the EXP003 model with respect to
    the reference $\Lambda$CDM cosmology, consistently with the
    general findings of \citet{Pisani_etal_2015}.  Quantitatively, the
    excess is around $50\%$ with a difference larger than $4\sigma$.
    The radius at which this excess starts to be significant decreases
    with redshift, being $R \sim 20, 15, 12$ \Mpch at $z=0, 0.55, 1$,
    respectively.}

\item{The shape of the stacked density profile
    (Fig.~\ref{fig:CDM_prof}) is qualitatively similar to what
    previously found in the literature (i.e density minima
    around the centres of the CV and overdense compensation regions at
    $R \sim R_{\rm eff}$).  The void profiles in cDE models are not
    significantly different from what observed in the standard
    cosmology. Nonetheless, we can observe that close to the CV
    centres, the EXP003 model generally displays a density $10 - 25
    \%$ smaller than $\Lambda$CDM, thus showing that CV tend to be
    {\em emptier} in cDE models. On the other hand, the compensative
    overdensity at $R \sim R_{\rm eff}$ in the EXP003 case looks more
    prominent than in $\Lambda$CDM. All of these features are expected
    considering that the evolution of the background perturbations in
    cDE scenarios is faster than $\Lambda$CDM due to the {\em fifth
      force} associated with the coupling.}
\end{enumerate}

In Section~\ref{voidChalo} we then focused on CV identified in the
halo distribution, finding that the comparison between cDE models and
the reference cosmology is very different from what found for CDM.
More specifically, we find the following results:

\begin{enumerate}

\item{The filling factor of CV in halos is not strongly dependent on
    the considered cDE model (Fig.~\ref{vol_frac}). Only minor, not
    significant differences are found in the volume fractions at all
    redshifts considered in this paper. This last result is starkly
    different from what observed in CV samples detected in the CDM
    distribution.}

\item{The comparison of DSD (Fig.~\ref{fig:halos_VSD}) in the halo
    distribution does not reveal sensible differences between cDE
    models and
    the reference one.  This result is again substantially different
    from what found in Section~\ref{CDM-CoDECS}. We connect this
    discrepancy with the impact of the halo bias on CV properties. To
    test such effect, we compare DSD of CV in a random subsample of
    the CDM distribution with the same density of tracers as the FoF
    halo catalog. Again, in this last case CV in CDM do not show the
    same relative trend in the DSD as for the CV in the halo
    distribution (Fig.~\ref{fig:CDM_VSD_CMB_halodens}).  The impact of
    the halo bias can be observed also by increasing the minimum mass
    of halos used as tracers: in Fig.~\ref{fig:bias_VSD} we show that
    including only halos with large masses ($> 5 \cdot 10^{12} \cdot
    M_{\bigodot}$) the cDE models show an excess of CV with $30 < R_{\rm
      eff}$[\Mpch]$ < 60$ at $z=0.55, 1$, which is not seen in
    Fig.~\ref{fig:halos_VSD}.}

\item{The density profile of CV in halos does not look like an
    effective probe to discriminate among cDE models. Indeed, as shown
    in Fig.~\ref{fig:halosprof}, the stacked profiles of CV in cDE
    models are only marginally distinguishable from the $\Lambda $CDM
    case, and only in the very innermost parts.}
\end{enumerate}

To conclude, the main result of this work is that the properties of CV
in different cosmological models are strongly affected by the choice
of the tracers of the underlying density field used to detect them
(halos or CDM particles): {this is caused by the impact of the halo
bias on the structural properties of CV: as the bias evolves differently for different cDE models,
this is reflected in a non-trivial way on the properties of the associated CV sample.}
Our results
indirectly challenge the assumption made in several recent works that
a subsampled distribution of simulated CDM particles with the same
density of the expected tracers of a real galaxy survey might provide
reliable predictions about the effective discriminating power of CV in
that survey.

\section*{Acknowledgments}

We are thankful to Gianni Zamorani for useful
hints on the statistical comparison between the models.
GP would like to thank Jochen Weller and Ben Hoyle for useful
suggestions and Paul Sutter for helpful discussions about {\small
  ZOBOV} and {\small VIDE}.  The numerical simulations presented in
this work have been performed and analysed on the Hydra cluster at the
RZG supercomputing centre in Garching. MB acknowledges support by
the Marie Curie Intra European Fellowship ``SIDUN'' within the 7th
Framework Programme of the European Commission. We acknowledge
financial contributions from contracts ASI/INAF I/023/12/0 and by the
PRIN MIUR 2010-2011``The dark Universe and the cosmic evolution of
baryons: from current surveys to Euclid''. MB and LM also
acknowledge the financial contribution by the PRIN INAF 2012 ``The
Universe in the box: multiscale simulations of cosmic structure''.

%*****************************************************************************
\bibliographystyle{mn2e}
\bibliography{bib,baldi_bibliography}

\begin{thebibliography}{77}
\expandafter\ifx\csname natexlab\endcsname\relax\def\natexlab#1{#1}\fi

\bibitem[{{Alcock} \& {Paczynski}(1979)}]{AlcockPaczynski1979}
{Alcock} C., {Paczynski} B., 1979, \nat, 281, 358

\bibitem[{Amendola(2000)}]{Amendola_2000}
Amendola L., 2000, Phys. Rev., D62, 043511

\bibitem[{Amendola(2004)}]{Amendola_2004}
Amendola L., 2004, Phys. Rev., D69, 103524

\bibitem[{{Amendola}(2004)}]{Amendola2004}
{Amendola} L., 2004, \prd, 69, 103524

\bibitem[{{Amendola} {et~al}\mbox{.}(2013){Amendola} {et~al.}}]{amendola2013}
{Amendola} L., {et~al.}, 2013, Living Reviews in Relativity, 16, 6

\bibitem[{{Baldi}(2011{\natexlab{a}})}]{Baldi_2011b}
{Baldi} M., 2011{\natexlab{a}}, \mnras, 414, 116

\bibitem[{{Baldi}(2011{\natexlab{b}})}]{Baldi_2011a}
{Baldi} M., 2011{\natexlab{b}}, \mnras, 411, 1077

\bibitem[{{Baldi}(2012{\natexlab{a}})}]{Baldi_2011c}
{Baldi} M., 2012{\natexlab{a}}, \mnras, 420, 430

\bibitem[{{Baldi}(2012{\natexlab{b}})}]{CoDECS}
{Baldi} M., 2012{\natexlab{b}}, \mnras, 422, 1028

\bibitem[{{Baldi}(2012{\natexlab{c}})}]{baldi2012}
{Baldi} M., 2012{\natexlab{c}}, \mnras, 422, 1028

\bibitem[{{Baldi} {et~al}\mbox{.}(2010){Baldi}, {Pettorino}, {Robbers}, \&
  {Springel}}]{Baldi_etal_2010}
{Baldi} M., {Pettorino} V., {Robbers} G., {Springel} V., 2010, \mnras, 403,
  1684

\bibitem[{Beynon {et~al}\mbox{.}(2012)Beynon, Baldi, Bacon, Koyama, \&
  Sabiu}]{Beynon_etal_2012}
Beynon E., Baldi M., Bacon D.~J., Koyama K., Sabiu C., 2012,
  Mon.Not.Roy.Astron.Soc., 422, 3546

\bibitem[{{Bos} {et~al}\mbox{.}(2012){Bos}, {van de Weygaert}, {Dolag}, \&
  {Pettorino}}]{bos2012}
{Bos} E.~G.~P., {van de Weygaert} R., {Dolag} K., {Pettorino} V., 2012, \mnras,
  426, 440

\bibitem[{{Boylan-Kolchin}, {Bullock} \& {Kaplinghat}(2011){Boylan-Kolchin},
  {Bullock}, \& {Kaplinghat}}]{Boylan2011}
{Boylan-Kolchin} M., {Bullock} J.~S., {Kaplinghat} M., 2011, \mnras, 415, L40

\bibitem[{{Brax} \& {Martin}(1999)}]{Brax1999}
{Brax} P.~H., {Martin} J., 1999, Physics Letters B, 468, 40

\bibitem[{{Bullock}(2010)}]{Bullok2010}
{Bullock} J.~S., 2010, ArXiv e-prints

\bibitem[{Carbone {et~al}\mbox{.}(2013)Carbone, Baldi, Pettorino, \&
  Baccigalupi}]{Carbone_etal_2013}
Carbone C., Baldi M., Pettorino V., Baccigalupi C., 2013, JCAP, 1309, 004

\bibitem[{{Carlesi} {et~al}\mbox{.}(2014{\natexlab{a}}){Carlesi}, {Knebe},
  {Lewis}, {Wales}, \& {Yepes}}]{Carlesi_etal_2014a}
{Carlesi} E., {Knebe} A., {Lewis} G.~F., {Wales} S., {Yepes} G.,
  2014{\natexlab{a}}, \mnras, 439, 2943

\bibitem[{{Carlesi} {et~al}\mbox{.}(2014{\natexlab{b}}){Carlesi}, {Knebe},
  {Lewis}, \& {Yepes}}]{Carlesi_etal_2014b}
{Carlesi} E., {Knebe} A., {Lewis} G.~F., {Yepes} G., 2014{\natexlab{b}},
  \mnras, 439, 2958

\bibitem[{{Cervantes} {et~al}\mbox{.}(2012){Cervantes}, {Marulli},
  {Moscardini}, {Baldi}, \& {Cimatti}}]{vera2012}
{Cervantes} V.~D.~V., {Marulli} F., {Moscardini} L., {Baldi} M., {Cimatti} A.,
  2012, ArXiv e-prints

\bibitem[{{Clampitt}, {Cai} \& {Li}(2013){Clampitt}, {Cai}, \&
  {Li}}]{clampitt2013}
{Clampitt} J., {Cai} Y.-C., {Li} B., 2013, \mnras, 431, 749

\bibitem[{{Clampitt} \& {Jain}(2014)}]{clampitt2014}
{Clampitt} J., {Jain} B., 2014, ArXiv e-prints

\bibitem[{{Colberg} {et~al}\mbox{.}(2005){Colberg}, {Sheth}, {Diaferio}, {Gao},
  \& {Yoshida}}]{colberg2005}
{Colberg} J.~M., {Sheth} R.~K., {Diaferio} A., {Gao} L., {Yoshida} N., 2005,
  \mnras, 360, 216

\bibitem[{Cui, Baldi \& Borgani(2012)Cui, Baldi, \&
  Borgani}]{Cui_Baldi_Borgani_2012}
Cui W., Baldi M., Borgani S., 2012, arXiv:1201.3568

\bibitem[{{D'Amico} {et~al}\mbox{.}(2011){D'Amico}, {Musso}, {Nore{\~n}a}, \&
  {Paranjape}}]{damico2011}
{D'Amico} G., {Musso} M., {Nore{\~n}a} J., {Paranjape} A., 2011, \prd, 83,
  023521

\bibitem[{{de Blok}(2010)}]{deBlok2010}
{de Blok} W.~J.~G., 2010, Advances in Astronomy, 2010, 5

\bibitem[{{Elyiv} {et~al}\mbox{.}(2015){Elyiv}, {Marulli}, {Pollina}, {Baldi},
  {Branchini}, {Cimatti}, \& {Moscardini}}]{elyiv2015}
{Elyiv} A., {Marulli} F., {Pollina} G., {Baldi} M., {Branchini} E., {Cimatti}
  A., {Moscardini} L., 2015, \mnras, 448, 642

\bibitem[{{Farrar} \& {Peebles}(2004)}]{Farrar2004}
{Farrar} G.~R., {Peebles} P.~J.~E., 2004, \apj, 604, 1

\bibitem[{{Finelli} {et~al}\mbox{.}(2014){Finelli}, {Garcia-Bellido}, {Kovacs},
  {Paci}, \& {Szapudi}}]{Finelli_etal_2014}
{Finelli} F., {Garcia-Bellido} J., {Kovacs} A., {Paci} F., {Szapudi} I., 2014,
  ArXiv e-prints

\bibitem[{{Gibbons} {et~al}\mbox{.}(2014){Gibbons}, {Werner}, {Yoshida}, \&
  {Chon}}]{gibbons2014}
{Gibbons} G.~W., {Werner} M.~C., {Yoshida} N., {Chon} S., 2014, \mnras, 438,
  1603

\bibitem[{Giocoli {et~al}\mbox{.}(2013)Giocoli, Marulli, Baldi, Moscardini, \&
  Metcalf}]{Giocoli_etal_2012}
Giocoli C., Marulli F., Baldi M., Moscardini L., Metcalf R.~B., 2013,
  arXiv:1301.3151

\bibitem[{Giocoli {et~al}\mbox{.}(2015)Giocoli, Metcalf, Baldi, Meneghetti,
  Moscardini, {et~al.}}]{Giocoli_etal_2015}
Giocoli C., Metcalf R.~B., Baldi M., Meneghetti M., Moscardini L., {et~al.},
  2015

\bibitem[{{Gregory}, {Thompson} \& {Tifft}(1978){Gregory}, {Thompson}, \&
  {Tifft}}]{GregoryThompson1978}
{Gregory} S.~A., {Thompson} L.~A., {Tifft} W.~G., 1978, in Bulletin of the
  American Astronomical Society, Vol.~10, Bulletin of the American Astronomical
  Society, p. 622

\bibitem[{{Hamaus}, {Sutter} \& {Wandelt}(2014){Hamaus}, {Sutter}, \&
  {Wandelt}}]{Hamaus2014}
{Hamaus} N., {Sutter} P.~M., {Wandelt} B.~D., 2014, Physical Review Letters,
  112, 251302

\bibitem[{{Hausman}, {Olson} \& {Roth}(1983){Hausman}, {Olson}, \&
  {Roth}}]{hausman1983}
{Hausman} M.~A., {Olson} D.~W., {Roth} B.~D., 1983, \apj, 270, 351

\bibitem[{{Izumi} {et~al}\mbox{.}(2013){Izumi}, {Hagiwara}, {Nakajima},
  {Kitamura}, \& {Asada}}]{izumi2013}
{Izumi} K., {Hagiwara} C., {Nakajima} K., {Kitamura} T., {Asada} H., 2013,
  \prd, 88, 024049

\bibitem[{{Kirshner} {et~al}\mbox{.}(1981){Kirshner}, {Oemler}, {Schechter}, \&
  {Shectman}}]{kirshner1981}
{Kirshner} R.~P., {Oemler}, Jr. A., {Schechter} P.~L., {Shectman} S.~A., 1981,
  \apjl, 248, L57

\bibitem[{{Kova{\v c}} {et~al}\mbox{.}(2014){Kova{\v c}}, {Lilly}, {Knobel},
  {Bschorr}, {Peng}, {Carollo}, {Contini}, {Kneib}, {Le F{\'e}vre}, {Mainieri},
  {Renzini}, {Scodeggio}, {Zamorani}, {Bardelli}, {Bolzonella}, {Bongiorno},
  {Caputi}, {Cucciati}, {de la Torre}, {de Ravel}, {Franzetti}, {Garilli},
  {Iovino}, {Kampczyk}, {Lamareille}, {Le Borgne}, {Le Brun}, {Maier},
  {Mignoli}, {Oesch}, {Pello}, {Montero}, {Presotto}, {Silverman}, {Tanaka},
  {Tasca}, {Tresse}, {Vergani}, {Zucca}, {Aussel}, {Koekemoer}, {Le Floc'h},
  {Moresco}, \& {Pozzetti}}]{kovac2014}
{Kova{\v c}} K. {et~al.}, 2014, \mnras, 438, 717

\bibitem[{{Krause} {et~al}\mbox{.}(2013){Krause}, {Chang}, {Dor{\'e}}, \&
  {Umetsu}}]{krause2013}
{Krause} E., {Chang} T.-C., {Dor{\'e}} O., {Umetsu} K., 2013, \apjl, 762, L20

\bibitem[{{Laureijs} {et~al}\mbox{.}(2011){Laureijs} {et~al.}}]{laureijs2011}
{Laureijs} R., {et~al.}, 2011, ArXiv e-prints

\bibitem[{{Lavaux} \& {Wandelt}(2010)}]{lavaux2010}
{Lavaux} G., {Wandelt} B.~D., 2010, \mnras, 403, 1392

\bibitem[{Lee \& {Baldi}(2011)}]{Lee_Baldi_2011}
Lee J., {Baldi} M., 2011, \apj ~in press, arXiv:1110.0015, \apj ~Submitted

\bibitem[{Li(2011)}]{Li_2011}
Li B., 2011, Mon.Not.Roy.Astron.Soc., 411, 2615

\bibitem[{Li \& Barrow(2011)}]{Li_Barrow_2011}
Li B., Barrow J.~D., 2011, Phys. Rev., D83, 024007

\bibitem[{{Li} \& {Zhao}(2009)}]{LiZhao2009}
{Li} B., {Zhao} H., 2009, \prd, 80, 044027

\bibitem[{Macci\`{o} {et~al}\mbox{.}(2004)Macci\`{o}, Quercellini, Mainini,
  Amendola, \& Bonometto}]{Maccio_etal_2004}
Macci\`{o} A.~V., Quercellini C., Mainini R., Amendola L., Bonometto S.~A.,
  2004, Phys. Rev., D69, 123516

\bibitem[{{Marulli}, {Baldi} \& {Moscardini}(2012){Marulli}, {Baldi}, \&
  {Moscardini}}]{Marulli_Baldi_Moscardini_2012}
{Marulli} F., {Baldi} M., {Moscardini} L., 2012, \mnras, 420, 2377

\bibitem[{{Melchior} {et~al}\mbox{.}(2014){Melchior}, {Sutter}, {Sheldon},
  {Krause}, \& {Wandelt}}]{melchior2014}
{Melchior} P., {Sutter} P.~M., {Sheldon} E.~S., {Krause} E., {Wandelt} B.~D.,
  2014, \mnras, 440, 2922

\bibitem[{{Moresco} {et~al}\mbox{.}(2014){Moresco}, {Marulli}, {Baldi},
  {Moscardini}, \& {Cimatti}}]{Moresco_etal_2014}
{Moresco} M., {Marulli} F., {Baldi} M., {Moscardini} L., {Cimatti} A., 2014,
  \mnras, 443, 2874

\bibitem[{{Nadathur} \& {Hotchkiss}(2015)}]{Nadathur2015}
{Nadathur} S., {Hotchkiss} S., 2015, ArXiv e-prints

\bibitem[{{Nadathur} {et~al}\mbox{.}(2015){Nadathur}, {Hotchkiss}, {Diego},
  {Iliev}, {Gottl{\"o}ber}, {Watson}, \& {Yepes}}]{Nadathur2014}
{Nadathur} S., {Hotchkiss} S., {Diego} J.~M., {Iliev} I.~T., {Gottl{\"o}ber}
  S., {Watson} W.~A., {Yepes} G., 2015, \mnras, 449, 3997

\bibitem[{{Neyrinck}(2008)}]{neyrinck2008}
{Neyrinck} M.~C., 2008, \mnras, 386, 2101

\bibitem[{{Nusser}, {Gubser} \& {Peebles}(2005){Nusser}, {Gubser}, \&
  {Peebles}}]{nusser2005}
{Nusser} A., {Gubser} S.~S., {Peebles} P.~J., 2005, \prd, 71, 083505

\bibitem[{{Odrzywo{\l}ek}(2009)}]{odrzywolek2009}
{Odrzywo{\l}ek} A., 2009, \prd, 80, 103515

\bibitem[{{Pace} {et~al}\mbox{.}(2015){Pace}, {Baldi}, {Moscardini}, {Bacon},
  \& {Crittenden}}]{Pace_etal_2015}
{Pace} F., {Baldi} M., {Moscardini} L., {Bacon} D., {Crittenden} R., 2015,
  \mnras, 447, 858

\bibitem[{{Peebles}(2001)}]{peebles2001}
{Peebles} P.~J.~E., 2001, \apj, 557, 495

\bibitem[{{Pettorino} \& {Baccigalupi}(2008)}]{Pettorino2008}
{Pettorino} V., {Baccigalupi} C., 2008, \prd, 77, 103003

\bibitem[{{Pisani} {et~al}\mbox{.}(2015){Pisani}, {Sutter}, {Hamaus},
  {Alizadeh}, {Biswas}, {Wandelt}, \& {Hirata}}]{Pisani_etal_2015}
{Pisani} A., {Sutter} P.~M., {Hamaus} N., {Alizadeh} E., {Biswas} R., {Wandelt}
  B.~D., {Hirata} C.~M., 2015, ArXiv e-prints

\bibitem[{{Planck Collaboration} {et~al}\mbox{.}(2015{\natexlab{a}}){Planck
  Collaboration}, {Adam}, {Ade}, {Aghanim}, {Akrami}, {Alves}, {Arnaud},
  {Arroja}, {Aumont}, {Baccigalupi}, \& et~al.}]{2015Planck}
{Planck Collaboration} {et~al.}, 2015{\natexlab{a}}, ArXiv e-prints

\bibitem[{{Planck Collaboration} {et~al}\mbox{.}(2015{\natexlab{b}}){Planck
  Collaboration}, {Ade}, {Aghanim}, {Arnaud}, {Ashdown}, {Aumont},
  {Baccigalupi}, {Banday}, {Barreiro}, {Bartlett}, \& et~al.}]{Planck2015param}
{Planck Collaboration} {et~al.}, 2015{\natexlab{b}}, ArXiv e-prints

\bibitem[{{Platen}, {van de Weygaert} \& {Jones}(2007){Platen}, {van de
  Weygaert}, \& {Jones}}]{Platen2007}
{Platen} E., {van de Weygaert} R., {Jones} B.~J.~T., 2007, \mnras, 380, 551

\bibitem[{{Ratra} \& {Peebles}(1988)}]{RatraPeebles1988}
{Ratra} B., {Peebles} P.~J.~E., 1988, \prd, 37, 3406

\bibitem[{{Rees}, {Sciama} \& {Stobbs}(1968){Rees}, {Sciama}, \&
  {Stobbs}}]{rees1968}
{Rees} M.~J., {Sciama} D.~W., {Stobbs} S.~H., 1968, \aplett, 2, 243

\bibitem[{{Ricciardelli}, {Quilis} \& {Planelles}(2013){Ricciardelli},
  {Quilis}, \& {Planelles}}]{ricciardelli2013}
{Ricciardelli} E., {Quilis} V., {Planelles} S., 2013, \mnras, 434, 1192

\bibitem[{{Ricciardelli}, {Quilis} \& {Varela}(2014){Ricciardelli}, {Quilis},
  \& {Varela}}]{ricciardelli2014}
{Ricciardelli} E., {Quilis} V., {Varela} J., 2014, \mnras, 440, 601

\bibitem[{{Spolyar}, {Sahl{\'e}n} \& {Silk}(2013){Spolyar}, {Sahl{\'e}n}, \&
  {Silk}}]{spolyar2013}
{Spolyar} D., {Sahl{\'e}n} M., {Silk} J., 2013, Physical Review Letters, 111,
  241103

\bibitem[{Springel(2005)}]{gadget-2}
Springel V., 2005, Mon. Not. Roy. Astron. Soc., 364, 1105

\bibitem[{{Sutter} {et~al}\mbox{.}(2015{\natexlab{a}}){Sutter}, {Carlesi},
  {Wandelt}, \& {Knebe}}]{sutter2014cDE}
{Sutter} P.~M., {Carlesi} E., {Wandelt} B.~D., {Knebe} A., 2015{\natexlab{a}},
  \mnras, 446, L1

\bibitem[{{Sutter} {et~al}\mbox{.}(2015{\natexlab{b}}){Sutter}, {Carlesi},
  {Wandelt}, \& {Knebe}}]{Sutter_etal_2015}
{Sutter} P.~M., {Carlesi} E., {Wandelt} B.~D., {Knebe} A., 2015{\natexlab{b}},
  \mnras, 446, L1

\bibitem[{{Sutter} {et~al}\mbox{.}(2015{\natexlab{c}}){Sutter}, {Lavaux},
  {Hamaus}, {Pisani}, {Wandelt}, {Warren}, {Villaescusa-Navarro}, {Zivick},
  {Mao}, \& {Thompson}}]{VIDE2015}
{Sutter} P.~M. {et~al.}, 2015{\natexlab{c}}, Astronomy and Computing, 9, 1

\bibitem[{{Sutter} {et~al}\mbox{.}(2012){Sutter}, {Lavaux}, {Wandelt}, \&
  {Weinberg}}]{sutter2012}
{Sutter} P.~M., {Lavaux} G., {Wandelt} B.~D., {Weinberg} D.~H., 2012, \apj,
  761, 187

\bibitem[{{Sutter} {et~al}\mbox{.}(2014){Sutter}, {Pisani}, {Wandelt}, \&
  {Weinberg}}]{sutter2014}
{Sutter} P.~M., {Pisani} A., {Wandelt} B.~D., {Weinberg} D.~H., 2014, \mnras,
  443, 2983

\bibitem[{{Szapudi} {et~al}\mbox{.}(2014){Szapudi}, {Kov{\'a}cs}, {Granett},
  {Frei}, {Silk}, {Garcia-Bellido}, {Burgett}, {Cole}, {Draper}, {Farrow},
  {Kaiser}, {Magnier}, {Metcalfe}, {Morgan}, {Price}, {Tonry}, \&
  {Wainscoat}}]{szapudi2014}
{Szapudi} I. {et~al.}, 2014, ArXiv e-prints

\bibitem[{{Tinker} \& {Conroy}(2009)}]{Tinker2009}
{Tinker} J.~L., {Conroy} C., 2009, \apj, 691, 633

\bibitem[{Weinberg(1989)}]{weinberg1989cosmological}
Weinberg S., 1989, Reviews of Modern Physics, 61, 1

\bibitem[{{Wetterich}(1988)}]{wetterich1988}
{Wetterich} C., 1988, Nuclear Physics B, 302, 668

\bibitem[{Wetterich(1995)}]{Wetterich_1995}
Wetterich C., 1995, Astron. Astrophys., 301, 321

\end{thebibliography}

\label{lastpage}

\end{document}